\documentclass[pre,aps,onecolumn,superscriptaddress,nofootinbib]{revtex4-1}
\pdfoutput=1
\usepackage{amssymb}
\usepackage{amsmath}
\usepackage{subfigure}
\usepackage{graphicx}
\usepackage{textcomp}
\usepackage{url}
\usepackage{color}
\usepackage{cases}
\usepackage[normalem]{ulem}

\usepackage[colorlinks=true, urlcolor=blue, anchorcolor=blue, citecolor=blue,filecolor=blue,linkcolor=blue,menucolor=blue
]{hyperref}

\usepackage[titletoc,title]{appendix}

\usepackage{color}

\begin{document}
\title{Time-averaged height distribution of the Kardar-Parisi-Zhang interface}

\author{Naftali R. Smith}
\email{naftali.smith@mail.huji.ac.il}
\author{Baruch Meerson}
\email{meerson@mail.huji.ac.il}
\author{Arkady Vilenkin}
\email{vilenkin@mail.huji.ac.il}
\affiliation{Racah Institute of Physics, Hebrew University of
Jerusalem, Jerusalem 91904, Israel}


\begin{abstract}

We study the complete probability distribution
$\mathcal{P}\left(\bar{H},t\right)$ of the time-averaged height $\bar{H}=(1/t)\int_0^t h(x=0,t')\,dt'$ at point $x=0$ of an evolving 1+1 dimensional Kardar-Parisi-Zhang (KPZ) interface $h\left(x,t\right)$. We focus on  short times  and flat initial condition and employ the optimal fluctuation method to determine the variance and the third cumulant of the distribution, as well as the asymmetric stretched-exponential tails. The tails scale as $-\ln\mathcal{P}\sim\left|\bar{H}\right|^{3/2} \! /\sqrt{t}$ and $-\ln\mathcal{P}\sim\left|\bar{H}\right|^{5/2} \! /\sqrt{t}$, similarly to
the previously determined tails of the one-point KPZ height statistics at specified time $t'=t$. The optimal interface histories, dominating these tails, are markedly different.  Remarkably, the optimal history, $h\left(x=0,t\right)$, of the interface height at $x=0$ is a non-monotonic function of time: the maximum (or minimum) interface height is achieved at an intermediate time. We also address a more general problem of determining  the probability density of observing a given height history of the KPZ interface at point $x=0$.

\end{abstract}

\maketitle


\tableofcontents
\nopagebreak

\section{Introduction}

Nonequilibrium stochastic surface growth continues to attract attention for more than three decades \cite{Barabasi,McKane,HHZ1,Krug}. The commonly used measures of surface growth are the interface width and the two-point spatial correlation function \cite{Barabasi}. When the interface is rough, these measures exhibit, at long times, dynamic scaling properties. Depending on the values of the corresponding exponents, the interfaces are divided into different universality classes \cite{Barabasi,McKane,HHZ1,Krug}. Although these measures provide a valuable insight, they
(and a more general measure -- the two-point correlation function both in space and in time \cite{Family1,Family2}) do not
fully capture such a complex object as a stochastically evolving interface. It is not surprising, therefore, that additional measures have been introduced. One group of such measures --  the persistence and first-passage properties of interfaces \cite{Krugetal,BMSreview} -- has been around since the nineties. More recently, the focus shifted towards more detailed quantities, such as the complete probability distribution $P\left(H,t\right)$ of the interface height $h\left(x=0,t\right)=H$ at
the origin at time $t$. This shift of focus was a result of a remarkable progress in the theory of this quantity for the 1+1 dimensional Kardar-Parisi-Zhang (KPZ) equation \citep{KPZ} -- see Eq.~(\ref{eq:KPZ_dimensional}) below -- which describes an important universality class of stochastic growth \citep{SS,CDR,Dotsenko,ACQ, CLD,IS,Borodinetal}. 

In this work we propose to characterize the interface height fluctuations by the probability distribution $\mathcal{P}\left(\bar{H},T\right)$ of the \emph{time-averaged} height at point $x=0$:
\begin{equation}\label{hbardefinition}
\bar{H}=\frac{1}{T} \int_0^T  h\left(x=0,t\right) \,dt .
\end{equation}
Fluctuation statistics of time-averaged quantities have attracted much recent interest in statistical mechanics, see \textit{e.g.} reviews  \cite{Derrida2007,Touchette2009,bertini2015,Touchette2018}. It is natural, therefore, to extend their use to a characterization of fluctuating interfaces.

A particular problem that we will consider here deals with the  KPZ equation that governs the evolution in time of the height of a growing stochastic interface $h\left(x,t\right)$ in 1+1 dimension:
\begin{equation}
\label{eq:KPZ_dimensional}
\partial_{t}h=\nu\partial_{x}^{2}h+\frac{\lambda}{2}\left(\partial_{x}h\right)^{2}+\sqrt{D}\,\xi(x,t).
\end{equation}
The interface is driven by  the Gaussian noise $\xi(x,t)$ which has zero average
and is white, that is uncorrelated, both in space and in time%
\footnote{We subtract from $h\left(x=0,t\right)$ in Eq.~(\ref{hbardefinition}) the systematic displacement of the interface that results from the rectification of the
noise by the KPZ nonlinearity \citep{S2016,Gueudre,Hairer}.}.
At late times, the KPZ interface width grows as $t^{1/3}$, and the lateral correlation length grows as $t^{2/3}$ \cite{KPZ}.
The exponents $2/3$ and $1/3$ have been traditionally viewed as the hallmarks of the KPZ universality class \cite{Barabasi,McKane,HHZ1,Krug}. The exact results \citep{SS,CDR,Dotsenko,ACQ, CLD,IS,Borodinetal} for the complete one-point height distribution $P\left(H,t\right)$ led to sharper criteria for the KPZ universality class, and to the discovery of universality subclasses based on the initial conditions,
see Refs. \citep{SS,CDR,Dotsenko,ACQ, CLD,IS,Borodinetal} and reviews \cite{S2016,Corwin,QS,HHT,Takeuchi2017} for details.

Traditionally and understandably, the focus of interest in the KPZ equation has been its long-time dynamic scaling properties.  More recently, interest arose in the \emph{short-time} fluctuations of the KPZ interface height at a point
\citep{KK2007,KK2008,KK2009,
Gueudre,MKV,DMRS,KMSparabola,Janas2016, KrajenbrinkLeDoussal2017, MeersonSchmidt2017, SMS2018, SKM2018, SmithMeerson2018,MV2018, KrajenbrinkLeDoussal2018, Asida2019}.
For flat initial condition, it takes time of order $t_{\lambda}=\nu^{5}/(D^{2}\lambda^{4})$ for the nonlinear term in Eq.~(\ref{eq:KPZ_dimensional}) to kick in. Therefore, at $t\ll t_{\lambda}$ typical fluctuations of the interface are Gaussian and described by the Edwards-Wilkinson equation \citep{EW1982}: Eq.~(\ref{eq:KPZ_dimensional}) with $\lambda=0$. Large deviations, however, ``feel" the presence of the KPZ nonlinearity from the start. The interest in the short-time dynamics
emerged as a part of a general interest in large deviations in the KPZ equation. It was amplified by the discovery \cite{KK2007,KK2009,MKV,DMRS} of a novel exponent in the stretched-exponential behavior of one of the two non-Gaussian tails of $P\left(H,t\right)$: the $\lambda H<0$ tail.  This tail appears already at $t>0$ and persists, at sufficiently large $H$, at all times. The important latter property was conjectured in Ref. \cite{MKV}, demonstrated in an explicit asymptotic calculation for  ``droplet" initial condition in Ref. \cite{SMP}, reproduced by other methods in Refs. \cite{KrajenbrinkLD2018tail,Corwinetal2018,Krajenbrinketal2018} and proved rigorously in Ref.  \cite{Tsai2018}.

Most of the previous works on the statistics of time-averaged quantities have considered the long-time limit, when the time is much longer than the characteristic relaxation time of the system to a steady state \cite{Derrida2007,Touchette2009,bertini2015,Touchette2018}. In such systems the long-time limit can be often described by the Donsker-Varadhan large deviation formalism \cite{DV,Olla}. The KPZ interface does not reach a steady state in a one-dimensional infinite system: it continues to roughen forever. As a result, the statistics of time-averaged height is not amenable to the Donsker-Varadhan formalism, and one should look for alternatives. One such alternative, based on a small parameter, is provided by the optimal fluctuation method (OFM), which is also known as the instanton method, the weak-noise theory,  and the macroscopic fluctuation theory. In this method the path integral of the stochastic process, conditioned on a given large deviation, is evaluated using the saddle-point approximation.
This leads to a variational problem, the solution of which gives the optimal (that is, most likely) path of the system, and the optimal realization of the noise.
The ``classical'' action, evaluated on the optimal path, gives the logarithm of the corresponding probability. The origin of the OFM is in condensed matter physics \cite{Halperin,Langer,Lifshitz,Lifshitz1988}, but the OFM was also applied in such diverse areas as  turbulence and turbulent transport
\citep{turb1,turb2,turb3}, diffusive lattice gases \citep{bertini2015},
stochastic reactions on lattices \citep{EK, MS2011}, \textit{etc.}
It was already applied in many works to the KPZ equation and related systems \citep{Mikhailov1991, GurarieMigdal1996,Fogedby1998, Fogedby1999,Nakao2003,KK2007,KK2008,KK2009,Fogedby2009,MKV,KMSparabola,Janas2016,
MeersonSchmidt2017, MSV_3d, SMS2018, SKM2018, SmithMeerson2018, MV2018, Asida2019}.

In this work we will employ the OFM to study fluctuations of the time-averaged height of the KPZ interface for the flat initial condition. We find that the short-time scaling behavior of the distribution $\mathcal{P}(\bar{H},t)$ is $-\ln\mathcal{P}\simeq S \left(\bar{H}\right)/\sqrt{t}$, the same as that of the one-point one-time statistics $P\left(H,t\right)$ \citep{KK2007,KK2009,MKV,DMRS}, but the details of these two problems are different. We determine the variance and the third cumulant of the distribution $\mathcal{P}(\bar{H},t)$ and the asymmetric stretched-exponential tails. The $\lambda H \to \infty$ tail behaves as
\begin{equation}\label{3/2tail}
-\ln \mathcal{P}(\bar{H},t) \simeq \frac{16\sqrt{2}\,\nu |\bar{H}|^{3/2}}{3 D |\lambda|^{1/2} t^{1/2}},
\end{equation}
whereas the $\lambda H \to -\infty$ tail is the following:
\begin{equation}
\label{5/2tail}
-\ln\mathcal{P}\left(\bar{H},t\right)\simeq
\frac{s_{0}\sqrt{\left|\lambda\right|}\,|\bar{H}|^{5/2}}{Dt^{1/2}},
\end{equation}
where $s_0\simeq 1.61$.  The tails exhibit the same scalings as the previously determined tails of the one-point one-time KPZ height distribution. The optimal interface histories, dominating these tails, are markedly different.  One non-intuitive finding is that the optimal history, $h\left(x=0,t\right)$, of the interface height at $x=0$, conditioned on a specified $\bar{H}$, is a non-monotonic function of time: it exhibits a maximum or minimum at an intermediate time.

The remainder of this paper is organized as follows. In Sec.~\ref{sec:OFM} we formulate the OFM's variational problem whose solution yields the large deviation function $S \left(\bar{H}\right)$. In Sec.~\ref{sec:cumulants} we derive the second and third cumulants of the distribution $\mathcal{P}(\bar{H},t)$. Sections \ref{sec:soliton} and \ref{sec:HD} deal with the $\lambda \bar{H} \to +\infty$ and $\lambda \bar{H} \to -\infty$ tails of the distribution.
In Sec. \ref{sec:general_soliton} we extend the technique of Sec. \ref{sec:soliton} to a more general problem of determining  the probability density of observing a given height history
\begin{equation}
\label{moregeneral}
h\left(x=0,t\right)=h_{0}\left(t\right)
\end{equation}
of the KPZ interface at $x=0$.
 We summarize and briefly discuss our results in Sec.~\ref{disc}. Some technical details are relegated to Appendices A and B.

\section{Optimal fluctuation method}

\label{sec:OFM}

It is convenient to rescale time by the averaging time $T$, $x$ by the diffusion length $\sqrt{\nu T}$, and the interface height $h$ by $\nu/|\lambda|$.
Now the KPZ equation~(\ref{eq:KPZ_dimensional}) takes the dimensionless form \citep{MKV}
\begin{equation}
\label{eq:KPZ_dimensionless}
\partial_{t}h=\partial_{x}^{2}h-\frac{1}{2}\left(\partial_{x}h\right)^{2}+\sqrt{\epsilon} \, \xi\left(x,t\right),
\end{equation}
where $\epsilon=D\lambda^{2}\sqrt{T}/\nu^{5/2}$ is the dimensionless noise magnitude, and we assume  without loss of generality%
\footnote{Changing the sign of $\lambda$ is  equivalent to changing the sign of $h$.}
that $\lambda<0$.
In the weak-noise (that is, short-time) limit, which formally corresponds to $\epsilon \to 0$,
the proper path integral of Eq.~(\ref{eq:KPZ_dimensionless}) can be evaluated by using the saddle-point approximation. This leads to a minimization problem for the action functional
\begin{equation}
\label{eq:sdyn_def}
s\left[h\left(x,t\right)\right]=\frac{1}{2}\int_{0}^{1}dt\int_{-\infty}^{\infty}dx\left[\partial_{t}h-\partial_{x}^{2}h+\frac{1}{2}\left(\partial_{x}h\right)^{2}\right]^{2}.
\end{equation}
The (rescaled) constraint (\ref{hbardefinition}),
\begin{equation}
\label{eq:hbar_def}
\int_{0}^{1}h\left(x=0,t\right)dt = \bar{H},
\end{equation}
can be incorporated by minimizing the modified action
\begin{equation}
\label{eq:stotal_def}
s_{\Lambda}=s\left[h\left(x,t\right)\right]-\Lambda\int_{0}^{1}dt\,h\left(x=0,t\right)
=s\left[h\left(x,t\right)\right]
-\Lambda\int_{0}^{1}dt\int_{-\infty}^{\infty}dx\delta\left(x\right)h\left(x,t\right),
\end{equation}
where $\Lambda$ is a Lagrange multiplier whose value is ultimately determined by $\bar{H}$.
As in the previous works \citep{Fogedby1998, KK2007, MKV}, we prefer to recast the ensuing Euler-Lagrange equation into Hamiltonian equations for the optimal history $h\left(x,t\right)$ of the height profile and its canonically conjugate ``momentum'' $\rho\left(x,t\right)$ which describes the optimal realization of the noise $\xi$:
\begin{eqnarray}
  \partial_{t} h &=& \delta \mathcal{H}/\delta \rho = \partial_{x}^2 h - \frac{1}{2} \left(\partial_x h\right)^2+\rho ,  \label{eqh}\\
  \partial_{t}\rho &=& - \delta \mathcal{H}/\delta h = - \partial_{x}^2 \rho - \partial_x \left(\rho \partial_x h\right) - \Lambda\delta\left(x\right),\label{eqrho}
\end{eqnarray}
where the Hamiltonian $\mathcal{H}$ is
\begin{equation}
\label{eq:Hamiltonian_def}
\mathcal{H}=\Lambda h\left(0,t\right)+\int_{-\infty}^{\infty}\!dx\,\rho\left[\partial_{x}^{2}h
-\frac{1}{2}\left(\partial_{x}h\right)^{2}+\rho/2\right] .
\end{equation}
The delta-function term in Eq.~(\ref{eqrho}) is specific to the constraint of the  time-averaged height: it comes from the second term in Eq.~(\ref{eq:stotal_def}) as we explain in Appendix \ref{appendix:OFMequations}. It describes an effective driving of the optimal noise $\rho\left(x,t\right)$ by a permanent point-like source.  The initial condition for the flat interface is
\begin{equation}
\label{eq:flat_IC}
h\left(x,t=0\right)=0.
\end{equation}
As the variation of $s_{\Lambda}$ must vanish at $t=1$, we obtain the boundary condition
\begin{equation}
\label{pT}
\rho\left(x,t=1\right)=0.
\end{equation}

After solving the OFM problem and returning from $\Lambda$ to $\bar{H}$, we can evaluate the rescaled action, which can be recast as
\begin{equation}
\label{eq:sdyn_recast}
s\left(\bar{H}\right)=\frac{1}{2}\int_{0}^{1}dt\int_{-\infty}^{\infty}dx\,\rho^{2}\left(x,t\right).
\end{equation}
This action is the short-time large-deviation function of the time-averaged height.
It gives $\mathcal{P}\left(\bar{H},T\right)$ up to pre-exponential factors: $-\ln \mathcal{P} \simeq s\left(\bar{H}\right)/\epsilon$, or
\begin{equation}
-\ln\mathcal{P}\left(\bar{H},T\right)\simeq\frac{\nu^{5/2}}{D\lambda^{2}\sqrt{T}}\;s\left(\frac{\left|\lambda\right|\bar{H}}{\nu}\right)
 \label{actiondgen}
\end{equation}
in the physical variables. The same scaling behavior was obtained for the one-point one-time statistics \citep{KK2007,KK2009,MKV,DMRS}, but the functions $s(\dots)$ are of course different.

\section{Lower cumulants}
\label{sec:cumulants}
At sufficiently small $\bar{H}$, or $\Lambda$, one can solve the OFM problem perturbatively in powers of $\bar{H}$ or $\Lambda$. Previously, expansion in powers of $\Lambda$ was used \citep{MKV,KrMe}. As we show below, it is advantageous
to switch to expansion in powers of $\bar{H}$  at some stage, as this enables one to evaluate the third cumulant of $\mathcal{P}$ by calculating an integral which only involves terms of leading order in $\bar{H}$.

\subsection{Second cumulant}

In the leading order, we can neglect the nonlinear terms in the OFM equations, yielding
\begin{eqnarray}
  \partial_t h &=& \partial_x^2 h +\rho ,\label{heqEW}\\
  \partial_t \rho &=& -\partial_x^2 \rho -\Lambda \delta(x). \label{peqEW}
\end{eqnarray}
These linear equations correspond to the OFM applied to the Edwards-Wilkinson (EW) equation \citep{EW1982}.

\begin{figure}
\includegraphics[width=0.32\textwidth,clip=]{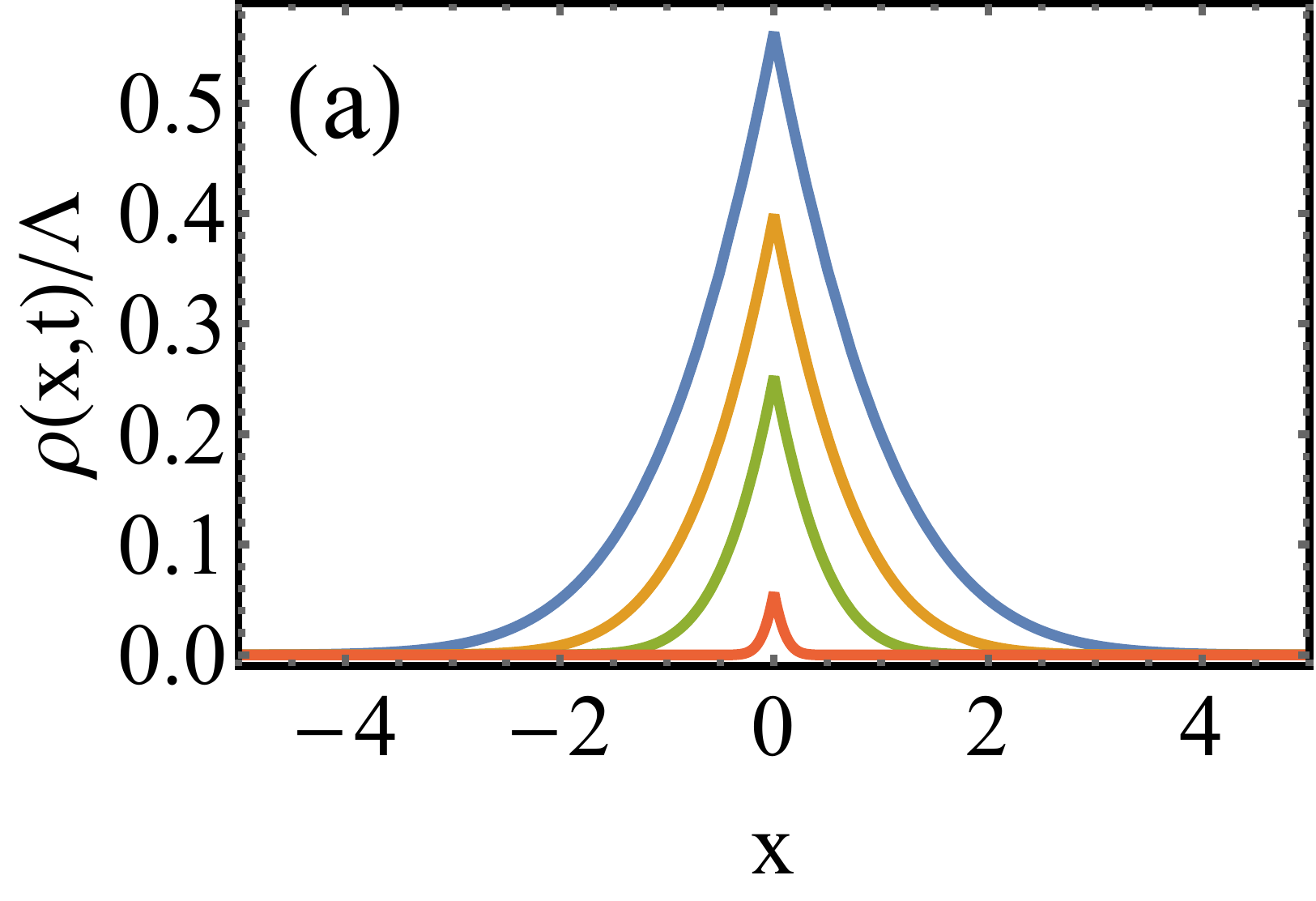}
\includegraphics[width=0.32\textwidth,clip=]{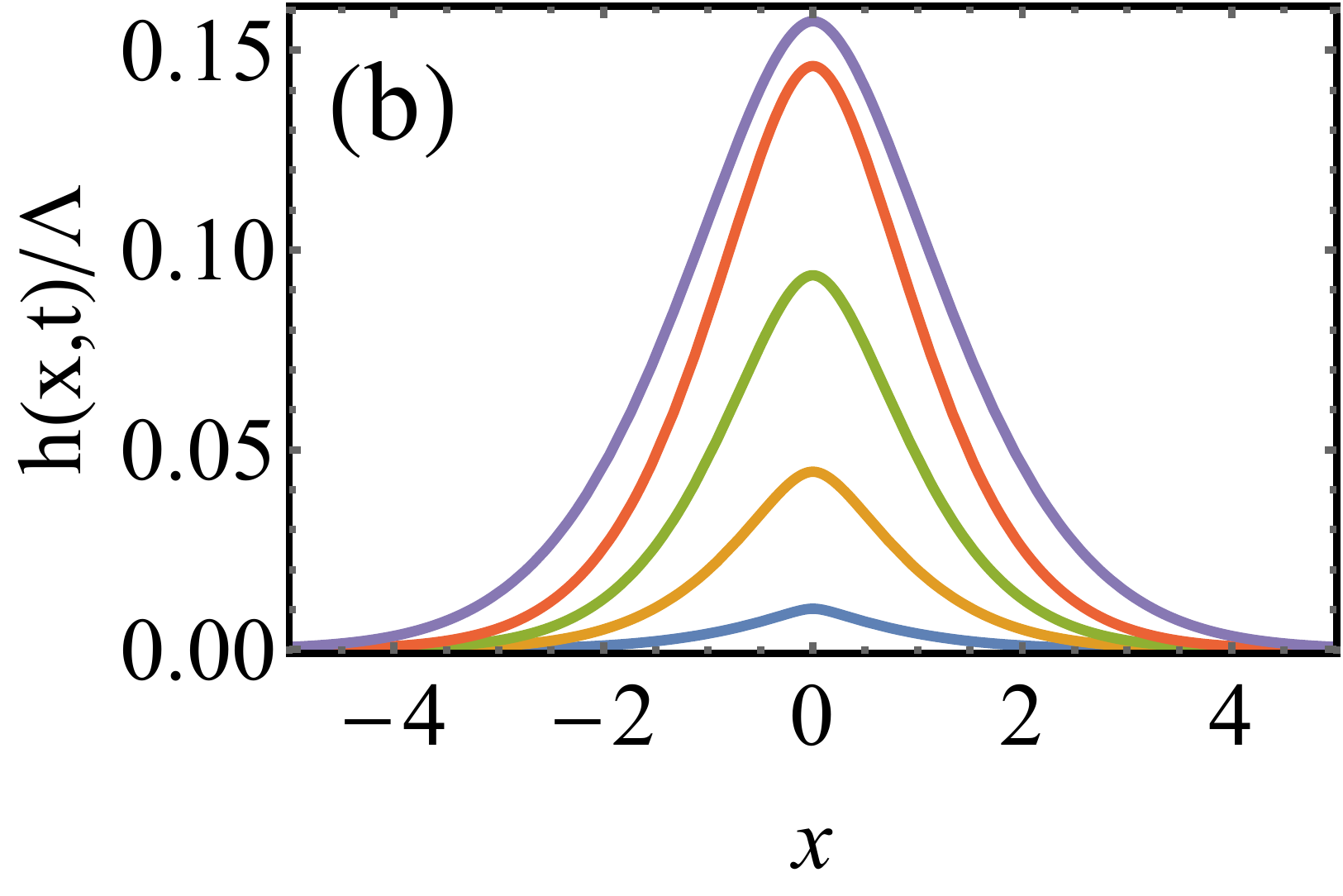}
\includegraphics[width=0.33\textwidth,clip=]{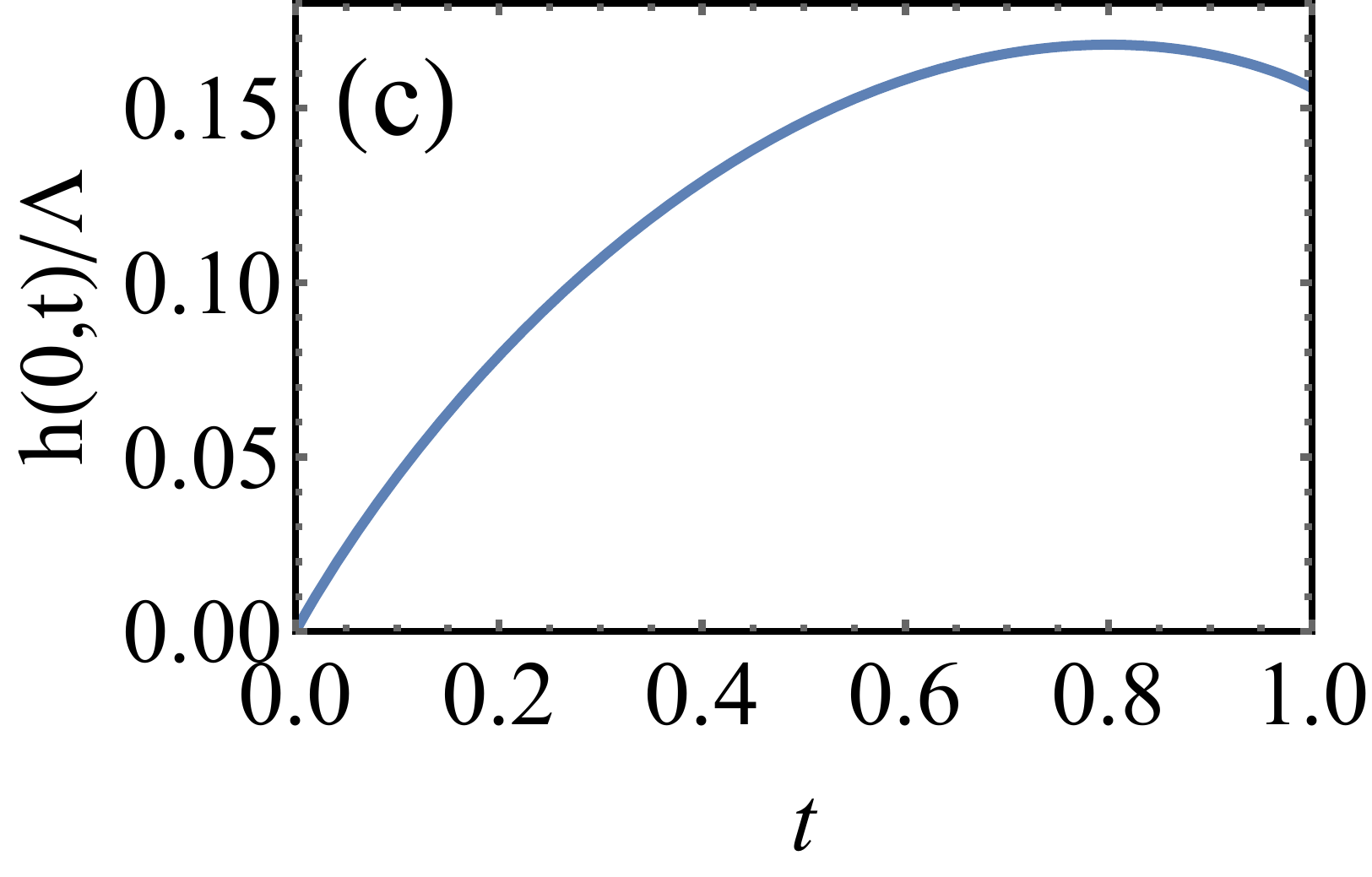}
\caption{(a) The optimal history of the noise $\rho\left(x,t\right)/\Lambda$ versus $x$ at rescaled times $t=0$, $0.5$, $0.8$ and $0.99$ (from top to bottom). (b) The optimal history of the interface at rescaled times $t=0.02$, $0.1$, $0.25$ and $0.99$ (from bottom to top). (c) $h\left(x=0,t\right)/\Lambda$ vs. $t$. The maximum is reached at $t=4/5$.}
\label{fig:EW}
\end{figure}

Solving Eq.~(\ref{peqEW}) backward in time with
the initial condition (\ref{pT}), we obtain
\begin{equation}\label{eq:rhoEW}
\rho\left(x,t\right)=\frac{\Lambda}{2}\left[x\,\text{erf}\left(\frac{x}{2\sqrt{1-t}}\right)-\left|x\right|+\frac{2\sqrt{1-t}}{\sqrt{\pi}}e^{-\frac{x^{2}}{4(1-t)}}\right]
\end{equation}
where $\text{erf}\left(z\right)$ is the error function. This solution is shown in Fig.~\ref{fig:EW} (a).
Now we can solve Eq.~(\ref{heqEW}) with the forcing term $\rho\left(x,t\right)$ from Eq. (\ref{eq:rhoEW}) and with the initial condition (\ref{eq:flat_IC}). The solution, shown in Fig.~\ref{fig:EW} (b),  is elementary but a bit cumbersome, see Appendix \ref{appendix:h_EW_regime}. At all times $t>0$, the maximum height is reached at $x=0$, as to be expected. Surprisingly, this maximum height,
\begin{equation}\label{hmaxvst}
h\left(x=0,t\right)=\frac{\Lambda\left[\left(1+t\right)^{3/2}-\left(1-t\right)^{3/2}-2t^{3/2}\right]}{3\sqrt{\pi}}
\end{equation}
is a non-monotonic function of time, see Fig.~\ref{fig:EW} (c). It reaches its maximum, $h_{\text{max}}=2 \Lambda/(3 \sqrt{5 \pi })$, at $t=4/5$.

Plugging Eq.~(\ref{eq:rhoEW}) into Eq.~(\ref{eq:sdyn_recast}), we find $s$ in terms of $\Lambda$:
\begin{equation}\label{svsLambda}
s=\frac{4 \left(\sqrt{2}-1\right) \Lambda ^2}{15 \sqrt{\pi }}.
\end{equation}
In its turn, plugging Eq.~(\ref{hmaxvst}) into~(\ref{eq:hbar_def}), we determine the relation between $\Lambda$ and  $\bar{H}$:
\begin{equation}
\label{hbarvsLambda}
\bar{H}=\frac{8\left(\sqrt{2}-1\right)\Lambda}{15\sqrt{\pi}}.
\end{equation}
Now Eqs.~(\ref{actiondgen}), (\ref{svsLambda}) and (\ref{hbarvsLambda})  yield, in the physical units, a Gaussian distribution%
\footnote{ \label{footnote_ds_dH_Lambda}
There is a simple exact relation $ds/d\bar{H}=\Lambda$ which is valid at all $\bar{H}$.
It can be proven explicitly, and it is a consequence of the fact that $\bar{H}$ and $\Lambda$ are conjugate variables.
A similar relation was quoted in Ref.~\cite{Chernykh}.
By virtue of this relation Eqs.~(\ref{svsLambda}) and~(\ref{hbarvsLambda}) can be derived from one another with no need for calculating $h\left(x,t\right)$.
}
\begin{equation}
\label{gauss}
-\ln\mathcal{P}\left(\bar{H},T\right)
\simeq\frac{15\sqrt{\pi\nu}\,\bar{H}^{2}}{16\left(\sqrt{2}-1\right)D\sqrt{T}}.
\end{equation}
The variance of this Gaussian distribution -- the second cumulant of the exact distribution $\mathcal{P}$ -- is
\begin{equation}\label{var}
\text{Var}_{\bar{H}}=\frac{8(\sqrt{2}-1)D \sqrt{T}}{15\sqrt{\pi \nu}}.
\end{equation}
It scales as $D\sqrt{T}/\nu$, as the second cumulant of the one-point one-time height distribution \cite{Krug1992,Gueudre,MKV},
but the numerical coefficient is different. Not surprisingly, it is smaller than in the one-point one-time problem.

\subsection{Third cumulant}
\label{sec:third_cumulant}

\begin{figure}
\includegraphics[width=0.4\textwidth,clip=]{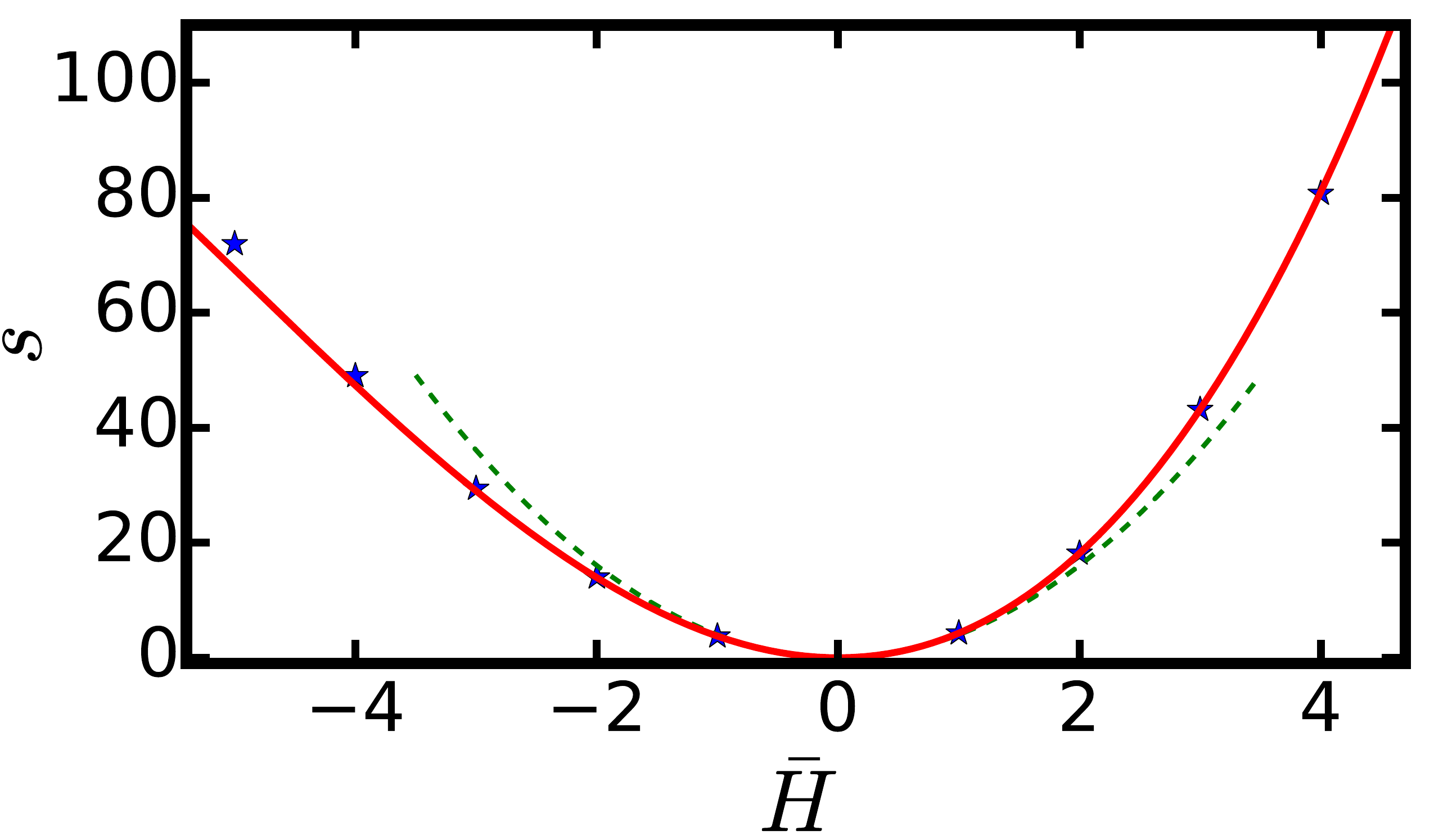}
\caption{Markers: direct computation of the action $s\left(\bar{H}\right)$ over numerical solutions to the OFM equations~(\ref{eqh}) and~(\ref{eqrho}). The dashed and solid lines are the quadratic and cubic approximations from Eq.~(\ref{cubic}), which we obtained analytically.}
\label{fig:third_cumulant}
\end{figure}

The third cumulant already ``feels" the KPZ nonlinearity. In order to calculate the third cumulant, it is advantageous to switch to the perturbative expansion in $\bar{H}$, rather than in $\Lambda$:
\begin{eqnarray}
\label{eq:h_perturbation}
h\left(x,t\right)&=&\bar{H}h_{1}\left(x,t\right)+\bar{H}^{2}h_{2}\left(x,t\right)+\dots,\\
\rho\left(x,t\right)&=&\bar{H}\rho_{1}\left(x,t\right)+\bar{H}^{2}\rho_{2}\left(x,t\right)+\dots.
 \end{eqnarray}
Correspondingly,
\begin{equation}\label{cubic}
s=\bar{H}^{2}s_{1}+\bar{H}^{3}s_{2}+\dots.
\end{equation}
Plugging Eq.~(\ref{eq:h_perturbation}) into the definition~(\ref{eq:sdyn_def}) of $s$ and taking leading order terms in $\bar{H}$ (see Appendix \ref{appendix:shortcut_3rd_cumulant}), we find
\begin{equation}
\label{eq:s2_shortcut}
s_{2}=\frac{1}{2}\int_{0}^{1}dt\int_{-\infty}^{\infty}dx\,\rho_{1}\left(\partial_{x}h_{1}\right)^{2}.
\end{equation}
Importantly, one does not have to determine the functions $h_2$ and $\rho_2$ in order to evaluate $s_{2}$. It suffices to know the quantities $h_1$ and $\rho_1$, determined in the previous iteration. Although they are known in analytical form, we were able to evaluate the double integral in Eq.~(\ref{eq:s2_shortcut}) only numerically. The result,
$s_{2} = 0.26308\dots$,  is in good agreement with a direct evaluation of $s$ over numerical solutions to the OFM
equations\footnote{In our numerical solutions of Eqs.~(\ref{eqh}) and (\ref{eqrho}) we used the Chernykh-Stepanov back-and-forth iteration algorithm \citep{Chernykh}.
},  see Fig.~\ref{fig:third_cumulant}.

With the cubic approximation (\ref{cubic}) for $s\left(\bar{H}\right)$ at hand,  it is straightforward to calculate the third cumulant of the distribution%
\footnote{See \textit{e.g.} Sec. 4.2.5 of the Supplemental Material in Ref. \citep{KrajenbrinkLeDoussal2017}.}:
\begin{equation}
\label{eq:kappa3}
\kappa_{3}=-\left.\frac{\partial_{\bar{H}}^{3}s}{\left(\partial_{\bar{H}}^{2}s\right)^{3}}\right|_{\bar{H}=0}\!\!\epsilon^{2}\left(\frac{\nu}{\lambda}\right)^{3}\,=\,
-\frac{3s_{2}\epsilon^{2}}{4s_{1}^{3}}\left(\frac{\nu}{\lambda}\right)^{3}
=0.003058\dots \,\frac{D^{2}\lambda T}{\nu^{2}}.
\end{equation}
The scaling $\kappa_3 \sim T$ is the same as of the third cumulant of the one-point one-time height distribution \cite{Gueudre,MKV}, but the numerical coefficient is different.

\section{$\lambda \bar{H} \to +\infty$ tail: Adiabatic soliton and everything around}
\label{sec:soliton}

\begin{figure}
\includegraphics[width=0.45\textwidth,clip=]{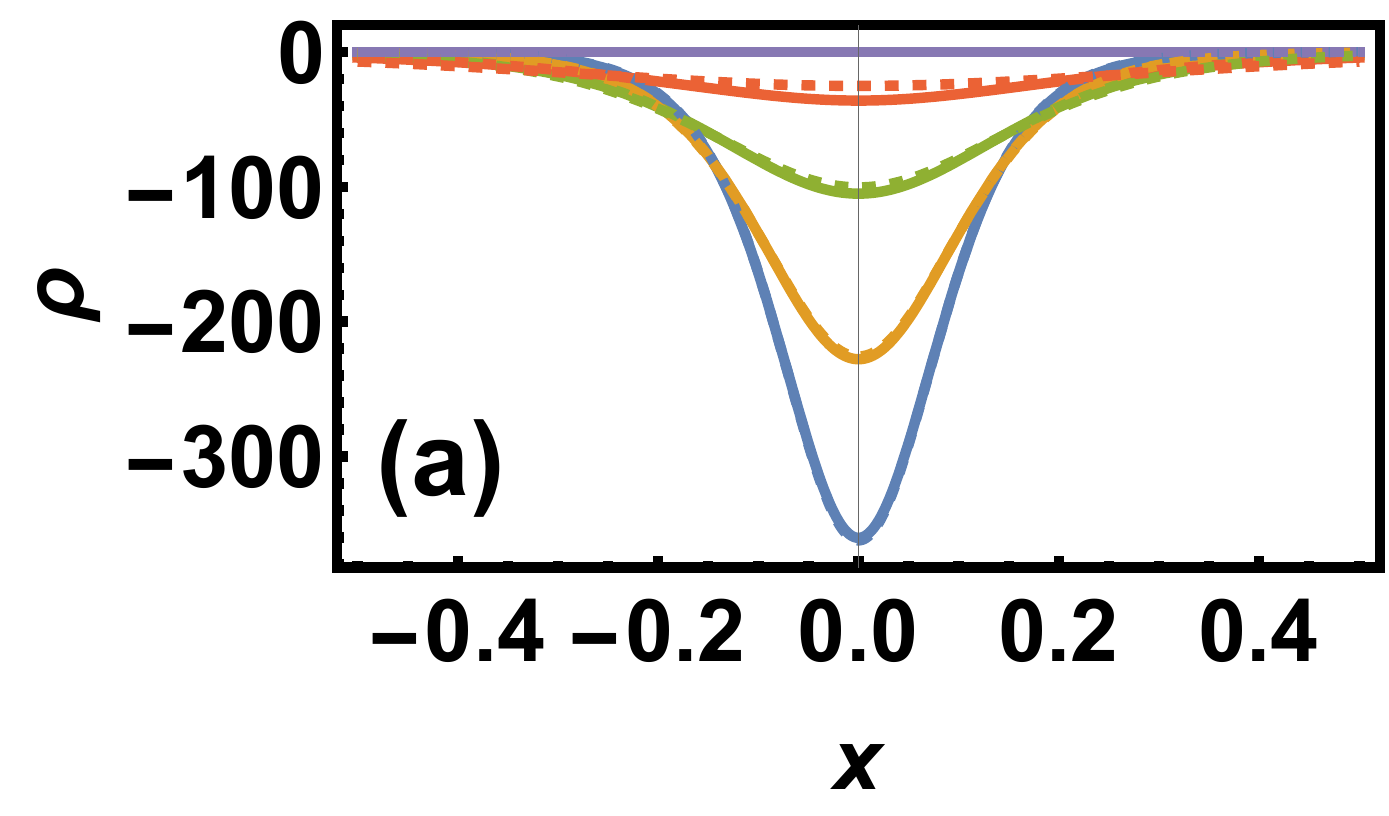}
\includegraphics[width=0.45\textwidth,clip=]{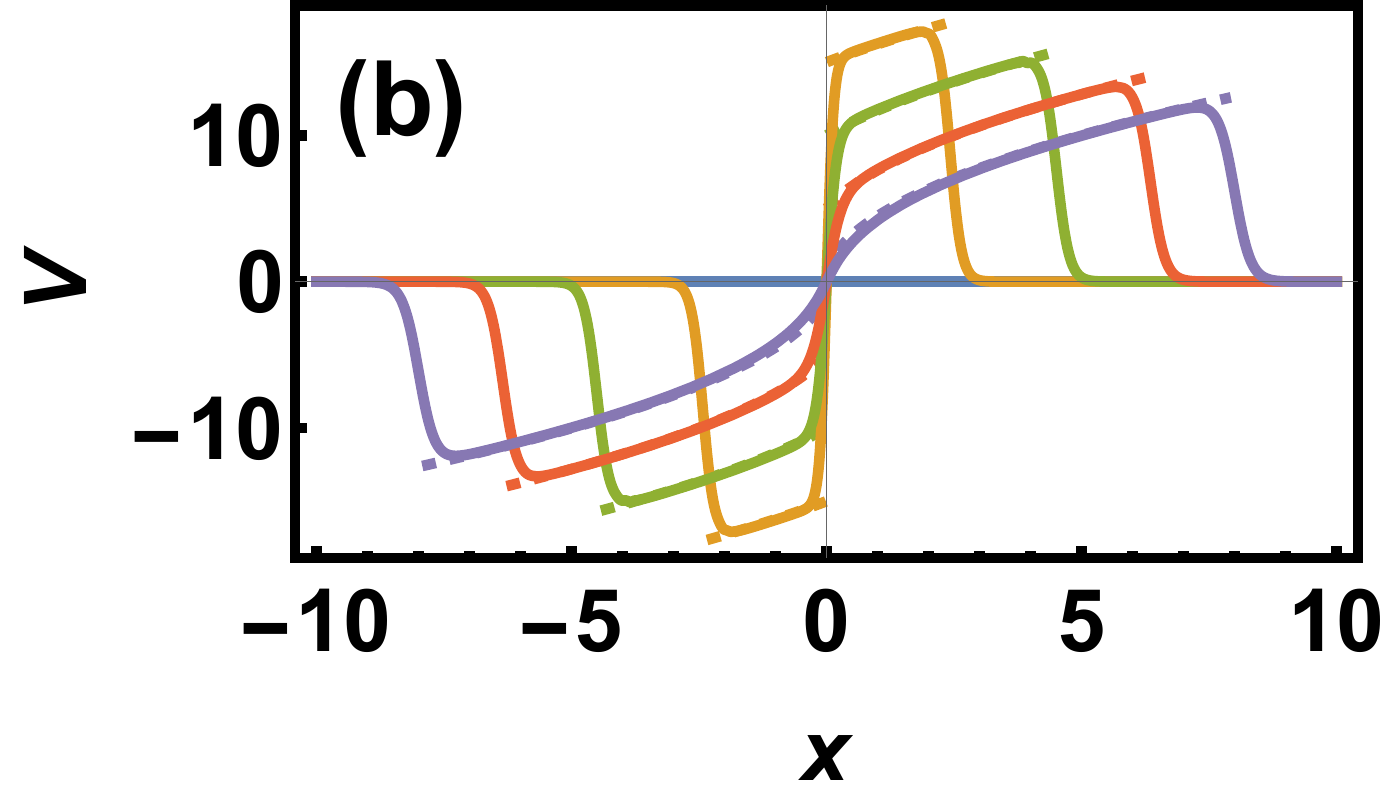}
\includegraphics[width=0.45\textwidth,clip=]{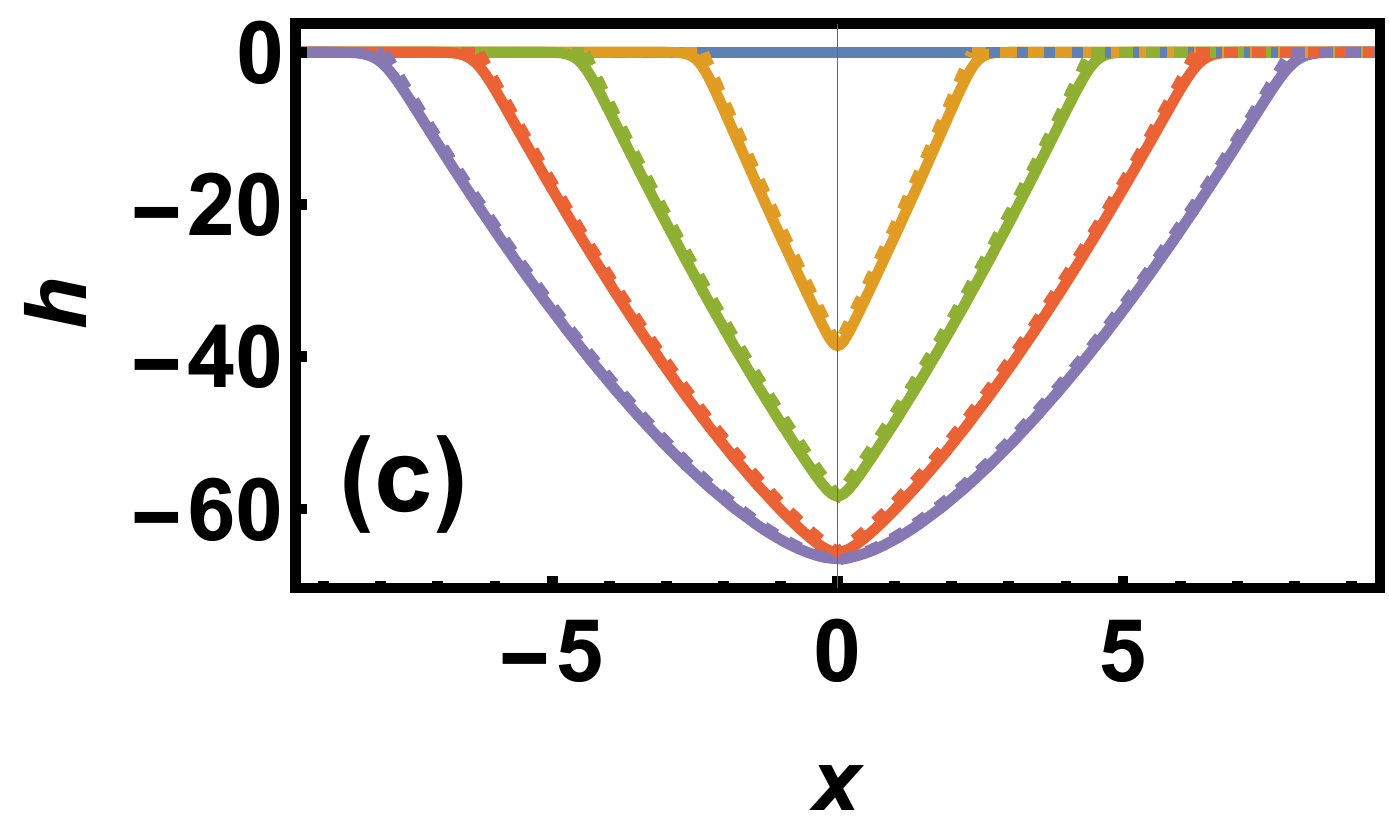}
\includegraphics[width=0.45\textwidth,clip=]{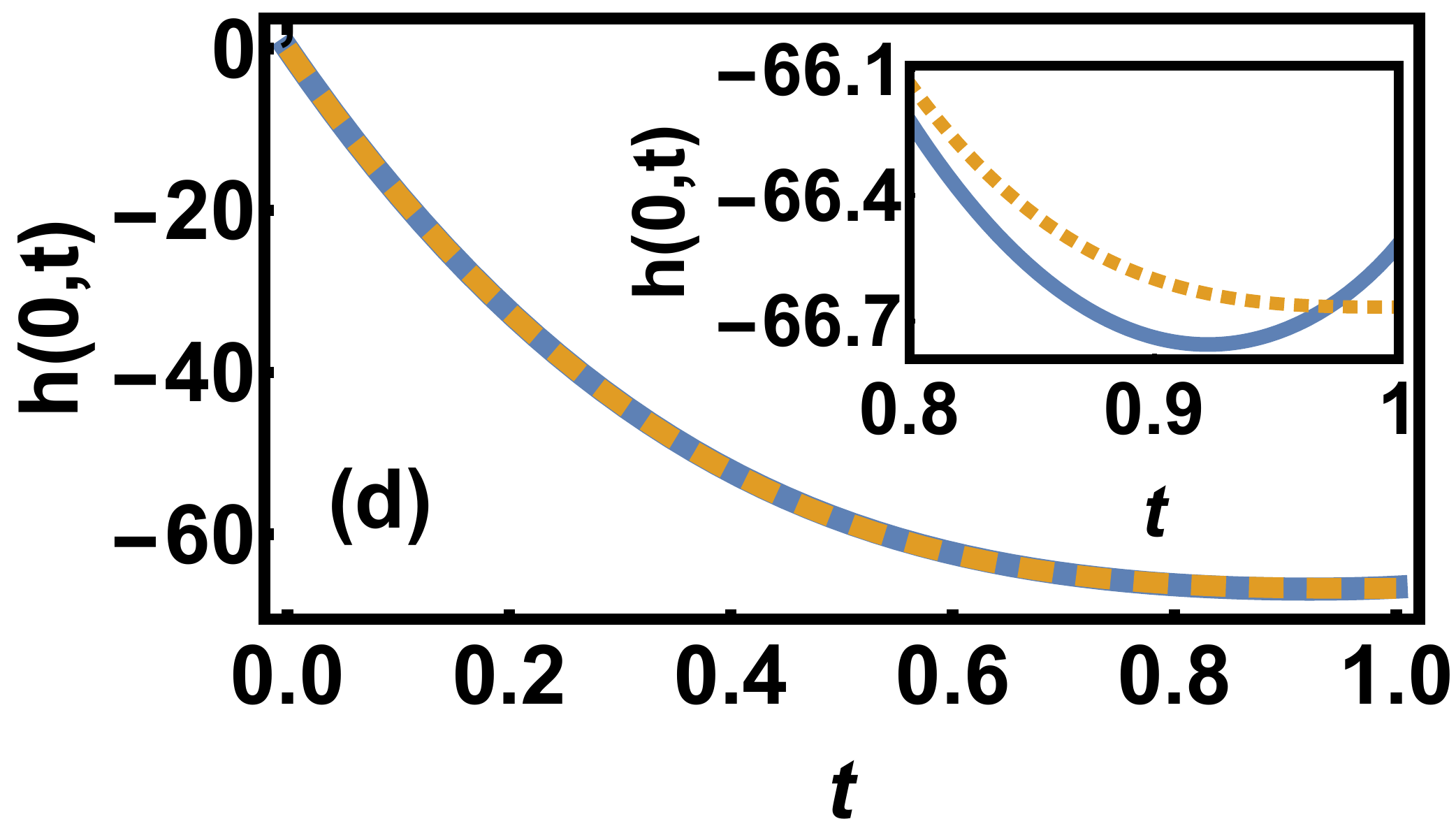}
\caption{Solid lines: numerical solutions
for the optimal path, corresponding to the $\lambda H\gg 1$ tail. Here $\bar{H} = -50$. Shown are (a) the adiabatically varying soliton $\rho\left(x,t\right)$ (bottom to top), (b) two adiabatically varying outgoing $V$-shocks $V\left(x,t\right)$, and (c) two adiabatically varying outgoing ``ramps" $h\left(x,t\right)$ at rescaled times $t=0,0.25,0.5,0.75,1$. Panel
(d) shows $h\left(x=0,t\right)$ vs.~time. The numerical results are almost indistinguishable from the analytical predictions~(\ref{eq:rho_adiabatic_soliton}),~(\ref{eq:c_of_t_for_time_average}),~(\ref{eq:h0_sol}),~(\ref{eq:Vsol_Hopf}) and~(\ref{eq:hsol_Hopf})  (dashed lines). The only exception is the inset of panel (d), but notice its vertical scale.  The horizontal scale in panel (a) is of the order of magnitude of the boundary layer, whereas in (b) and (c) it is of the order of magnitude of the  Hopf length scale $\sim \sqrt{-\bar{H}}$.}
\label{fig:negative_tail}
\end{figure}

Our numerics
show that, at $-\bar{H} \gg 1$, $\rho\left(x,t\right)$ is exponentially localized within a narrow boundary layer around $x=0$, see Fig.~\ref{fig:negative_tail} (a). A similar feature is present in the $-H \gg 1$ tail of the one-point one-time distribution for deterministic initial conditions \cite{KK2007,MKV,KMSparabola}. We now present an analytic theory in this limit, based on matched asymptotic expansions.  The idea is to approximately solve the OFM equations  Eqs.~(\ref{eqh}) and~(\ref{eqrho}) in an inner region which includes the boundary layer around $x=0$ and match this solution to the ``outer'' solution where both $\rho$ and the diffusion term in Eq.~(\ref{eqh}) are negligibly small. The outer region is described by the Hopf equation
\begin{equation}
\label{eq:Hopf}
\partial_tV +V\partial_{x}V=0
\end{equation}
for the interface slope field $V\left(x,t\right)= \partial_x h\left(x,t\right)$, and it can be divided into two sub-regions. At $\left|x\right|>x_{s}\left(t\right)$, where $\pm x_{s}\left(t\right)$ are the positions of two $V$-shocks determined below, $h$ (and therefore $V$) are negligibly small due to the flat initial condition, whereas in the region $\delta\left(t\right)\ll\left|x\right|<x_{s}\left(t\right)$ [where $\delta\left(t\right)$ denotes the width of the boundary layer and is found below] the Hopf flow is nontrivial and more complicated than the analogous Hopf flows in one-point one-time problems considered previously \citep{MKV, KMSparabola}. The whole outer region does not contribute to the action in the leading order. We begin by presenting the solution in the inner region and then we use it in order to evaluate the action. Afterwards, we find the outer solution and then match the two solutions in their joint region of validity.

In the one-point one-time problem \cite{KK2007,MKV}, one encounters an important class of exact solutions to the OFM equations [without the driving term $-\Lambda\delta\left(x\right)$ in Eq.~(\ref{eqrho})], describing a strongly localized soliton of $\rho$ and two outgoing ``ramps'' of $h$ \cite{Mikhailov1991,Fogedby1999}:
\begin{eqnarray}
\rho\left(x,t\right)&=&-2c\,\text{sech}^{2}\left(\sqrt{\frac{c}{2}}\,x\right),\\
\label{eq:h_exact_soliton}
h\left(x,t\right)&=&2\ln\cosh\left(\sqrt{\frac{c}{2}}\,x\right)-ct,
\end{eqnarray}
where $c>0$ is the velocity of the $h$-front in the vertical direction.
In the present problem the driving term in Eq.~(\ref{eqrho}) causes the soliton-front solution to vary with time adiabatically:
\begin{eqnarray}
\label{eq:rho_adiabatic_soliton}
\rho\left(x,t\right)&=&-2c\left(t\right)\text{sech}^{2}\left[\sqrt{\frac{c\left(t\right)}{2}}\,x\right],\\
\label{eq:h_adiabatic_soliton}
h\left(x,t\right)&=&2\ln\cosh\left[\sqrt{\frac{c\left(t\right)}{2}}\,x\right]+h_{0}\left(t\right),
\end{eqnarray}
where $h_{0}\left(t\right) = h\left(x=0,t\right)$ is the interface height at the origin, and the function $c\left(t\right)$ to be found is much larger than $1$.
The function $c\left(t\right)$ is related to $\Lambda$ through the balance equation
\begin{equation}
\label{eq:rho_conservation_with_Lambda}
\frac{d}{dt}\int\rho dx=-\Lambda,
\end{equation}
which can be obtained by integrating the both parts of Eq.~(\ref{eqrho}) over $x$ from $-\infty$ to $\infty$.
Plugging Eq.~(\ref{eq:rho_adiabatic_soliton}) into~(\ref{eq:rho_conservation_with_Lambda}), we obtain
\begin{equation}
\label{eq:Lambda0_and_c}
\frac{2\sqrt{2}}{\sqrt{c\left(t\right)}}\frac{dc}{dt} = \Lambda.
 \end{equation}
Furthermore, we can find a connection between $c\left(t\right)$ and $h_{0}\left(t\right)$.
Plugging the ansatz~(\ref{eq:rho_adiabatic_soliton}) and~(\ref{eq:h_adiabatic_soliton}) into Eq.~(\ref{eqh}) we obtain
\begin{equation}
\label{eq:c_h0dot_with_extra_term}
\frac{x}{\sqrt{c\left(t\right)}}\tanh\left[\sqrt{\frac{c\left(t\right)}{2}}\,x\right]\frac{dc}{dt}+\frac{dh_{0}\left(t\right)}{dt}=-c\left(t\right).
 \end{equation}
We now identify the inner region as $|x|\ll \Delta\left(t\right)$, where
\begin{equation}
\label{eq:boundary_layer_validity}
\Delta\left(t\right)=\frac{\left[c\left(t\right)\right]^{3/2}}{\dot{c}\left(t\right)}.
\end{equation}
[As one can check \textit{a posteriori}, $\Delta\left(t\right)$ is much larger than the characteristic soliton width
$\delta\left(t\right)=1/\sqrt{c(t)}$.] In the region $|x|\ll \Delta\left(t\right)$ the first term in Eq.~(\ref{eq:c_h0dot_with_extra_term}) is negligible, and we obtain the expected adiabatic relation
\begin{equation}
\label{eq:c_h0dot}
\frac{dh_{0}(t)}{dt} = -c\left(t\right).
 \end{equation}
We now integrate the ordinary differential equations~(\ref{eq:Lambda0_and_c}) and~(\ref{eq:c_h0dot}) subject to $h_{0}\left(t=0\right)=0$, Eq.~(\ref{eq:hbar_def}), and $c\left(t=1\right)=0$ [the latter condition follows from the boundary condition~(\ref{pT}) and Eq.~(\ref{eq:rho_adiabatic_soliton})]%
\footnote{ \label{footnote:adiabatic_soliton}
The adiabatic soliton-front ansatz~(\ref{eq:rho_adiabatic_soliton}) and~(\ref{eq:h_adiabatic_soliton}) is only valid for $c\gg1$. There are short temporal boundary layers near $t=0$ and $t=1$ where the solution rapidly adapts to the boundary conditions~(\ref{eq:flat_IC}) and~(\ref{pT}). In the limit $-\bar{H} \gg 1$ the relative contribution of these temporal boundary layers to the action is negligible.
}.
We obtain $\bar{H}=-\Lambda^{2}/128$, and
\begin{eqnarray}
\label{eq:c_of_t_for_time_average}
c\left(t\right) &=& 4\left|\bar{H}\right|\left(1-t\right)^{2},\\
\label{eq:h0_sol}
h_{0}\left(t\right)&=&\frac{4}{3}\bar{H}\left[1-\left(1-t\right)^{3}\right].
\end{eqnarray}
Note that the leading-order asymptotic result (\ref{eq:h0_sol}) describes a monotonically decreasing function of $t$. Numerics show that $h_{0}\left(t\right)$ is in fact not monotonic, but the local minimum at an intermediate time is very shallow, see Fig.~\ref{fig:negative_tail} (d). The non-monotonicity should appear in a subleading order of the theory with respect to $-\bar{H}\gg 1$.

Now we can evaluate the action. Plugging Eq.~(\ref{eq:rho_adiabatic_soliton}) into Eq.~(\ref{eq:sdyn_recast}) and using Eq.~(\ref{eq:c_of_t_for_time_average}), we obtain
\begin{equation}
\label{eq:action_negative_tail}
 s=\frac{1}{2}\int_{0}^{1}4c^{2}\left(t\right)dt\int_{-\infty}^{\infty}dx\,\text{sech}^{4}\left[\sqrt{\frac{c\left(t\right)}{2}}\,x\right]
 =\frac{8\sqrt{2}}{3}\int_{0}^{1} \! \left[c\left(t\right)\right]^{3/2}dt=
\frac{16\sqrt{2}}{3}\left|\bar{H}\right|^{3/2}
\end{equation}
which leads,  in the physical units, to the announced equation~(\ref{3/2tail}).
This tail scales as the previously determined $\lambda H \to +\infty$ tail of the one-point one-time distribution $P\left(H,T\right)$ \cite{KK2007,MKV}, but the coefficient in the exponent of $\mathcal{P}\left(\bar{H},T\right)$ is twice as large.

Using Eq.~(\ref{eq:c_of_t_for_time_average}), we can evaluate the characteristic soliton width, $\delta\left(t\right) \sim |\bar{H}|^{-1/2} (1-t)^{-1}$. At $-\bar{H}\gg 1$, $\delta\left(t\right)$ is much smaller than $1$ until very close to $t=1$.  In its turn, the length scale $\Delta\left(t\right) \sim |\bar{H}|^{1/2}(1-t)^2$ is much larger than $1$, except very close to $t=1$. Finally, the condition $c\left(t\right)\gg 1$ also holds except very close to $t=1$.

As it is evident from the calculations in Eq.~(\ref{eq:action_negative_tail}), the large deviation function $s\left(\bar{H}\right)$ comes only from the adiabatically varying $\rho$-soliton, which is exponentially localized within the boundary layer of width $\delta\left(t\right)$ around $x=0$. Still, for completeness,  we now determine the optimal path $h\left(x,t\right)$ in the outer region, that is outside this boundary layer. Here $\rho$ is negligible, and we can also neglect the diffusion term in Eq.~(\ref{eqh}) which, similarly to Ref.~\cite{MKV}, bring us to the Hopf equation (\ref{eq:Hopf}).
We will now solve this equation and match the solution with the inner solution in their joint region of their validity.  It suffices to solve for $x>0$. The solution for $x<0$ is obtained from the mirror symmetry of the problem around $x=0$.

The general solution to Eq.~(\ref{eq:Hopf}) is given in the implicit form by \citep{LL}
\begin{equation}
\label{eq:Hopf_gen_sol}
F\left(V\right)=x-Vt,
\end{equation}
where $F\left(V\right)$ is an arbitrary function. The joint region of validity of the Hopf solution and the boundary-layer solution is $\delta\left(t\right)\ll x\ll\Delta\left(t\right)$.
Using Eq.~(\ref{eq:c_of_t_for_time_average}), we find that the joint region is%
\footnote{See footnote \ref{footnote:adiabatic_soliton}.}
\begin{equation}
\label{eq:joint_region}
\frac{1}{\left|\bar{H}\right|^{1/2}\left(1-t\right)}\ll x\ll\left|\bar{H}\right|^{1/2}\left(1-t\right)^{2}.
\end{equation}
In this region Eqs.~(\ref{eq:h_adiabatic_soliton}) and~(\ref{eq:c_of_t_for_time_average}) lead to
\begin{equation}
\label{eq:V_joint_region}
V\simeq\sqrt{2c\left(t\right)}=2\sqrt{-2\bar{H}}\,\left(1-t\right),
\end{equation}
which is easily inverted
\begin{equation}
\label{eq:Hopf_matching_condition}
t\simeq1-\frac{V}{2\sqrt{-2\bar{H}}}.
\end{equation}
We now match Eq.~(\ref{eq:Hopf_matching_condition}) with the Hopf solution in the joint region~(\ref{eq:joint_region}). As we shall check later, in this region $x$ is negligible compared with the other terms in Eq.~(\ref{eq:Hopf_gen_sol}), so Eqs.~(\ref{eq:Hopf_gen_sol}) and~(\ref{eq:Hopf_matching_condition}) yield
\begin{equation}
\label{eq:Fsol}
F\left(V\right)=-V+\frac{V^2}{2\sqrt{-2\bar{H}}}.
\end{equation}
With $F\left(V\right)$ at hand, we can obtain the solution $V\left(x,t\right)$ in the entire nontrivial Hopf region
\begin{equation}
\label{eq:nontrivial_Hopf_region}
\delta\left(t\right)\ll x<x_{s}\left(t\right),
\end{equation}
where the $V$-shock position $x_{s}\left(t\right)$ is determined below. Plugging Eq.~(\ref{eq:Fsol}) back into~(\ref{eq:Hopf_gen_sol}) and solving for $V$,
we obtain a self-similar expression
\begin{equation}
\label{eq:Vsol_Hopf}
V\left(x,t\right)=\sqrt{-2\bar{H}}\,\left(1-t\right)\,\Phi\left[\frac{\sqrt{2} x}{\sqrt{-\bar{H}}(1-t)^2}\right],
\end{equation}
where $\Phi(z)=1+\sqrt{1+z}$,
and we chose the plus sign in front of $\sqrt{1+z}$ so that $V>0$ at $x>0$, corresponding to a minimum of $h$ at $x=0$.
One can now verify that in the joint region~(\ref{eq:joint_region}) $x$ is negligible in Eq.~(\ref{eq:Vsol_Hopf}) so that the latter equation indeed reduces to~(\ref{eq:V_joint_region}).
It is now straightforward to find the optimal history of the interface height $h\left(x,t\right)$ in the nontrivial Hopf region~(\ref{eq:nontrivial_Hopf_region}) from the relation
\begin{equation}\label{hsimple}
h\left(x,t\right)= h\left(0,t\right)+\int_0^x V(x',t) dx'.
\end{equation}
With Eqs.~(\ref{eq:h0_sol}) and (\ref{eq:Vsol_Hopf}), this yields
\begin{numcases}
{\frac{h\left(x,t\right)}{\bar{H}}\simeq}
\frac{4}{3}-(1-t)^3 f\left[\frac{\sqrt{2} |x|}{\sqrt{-\bar{H}}(1-t)^2}\right],   & $|x|<x_s\left(t\right)$, \label{eq:hsol_Hopf}
\\
0,   & $|x|>x_s\left(t\right)$,
\end{numcases}
where
\begin{equation}\label{f}
f\left(z\right)=\int_{0}^{z}\Phi\left(z'\right)dz'+\frac{4}{3}=\frac{2}{3}\left(1+z\right)^{3/2}+\frac{2}{3}+z,
\end{equation}
and we symmetrized the solution around $x=0$.

\begin{figure}
\includegraphics[width=0.45\textwidth,clip=]{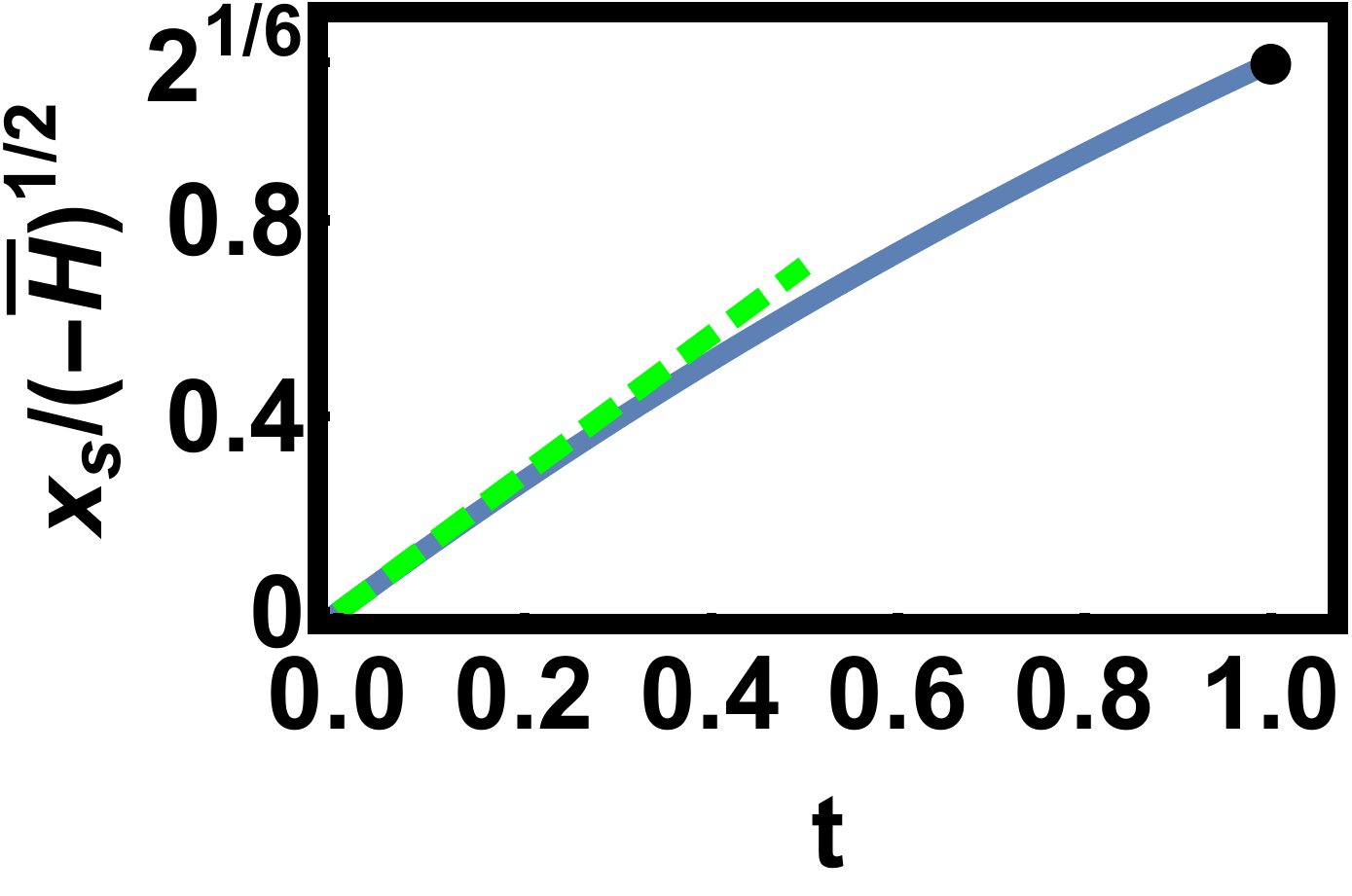}
\caption{Solid line: The location $x_{s}\left(t\right)$ of the $V$-shock which separates the nontrivial Hopf region~(\ref{eq:nontrivial_Hopf_region}) from the trivial region where $\rho=h=V=0$ in the $-\bar{H} \gg 1$ tail, see  Eqs.~(\ref{eq:hsol_Hopf}) and (\ref{xs}). The shock velocity $\dot{x}_{s}\left(t\right)$ goes down with time.
Dashed line: the asymptote $x_s\left(t \ll 1\right)\simeq \sqrt{-2\bar{H}} \, t$.
The fat dot denotes the point $x_s\left(t=1\right)=2^{1/6} \sqrt{-\bar{H}}$.
}
\label{fig:x_shock}
\end{figure}
The $V$-shock positions $\pm x_s\left(t\right)$ are described by the equation
\begin{equation}\label{xs}
x_{s}\left(t\right)=\sqrt{-\frac{\bar{H}}{2}}\,\left(1-t\right)^{2}\,f^{-1}\left[\frac{4}{3\left(1-t\right)^{3}}\right],
\end{equation}
where $f^{-1}(z)$ is the inverse of the function $f(z)$.
We do not give here the rather cumbersome expression for $f^{-1}$, but the resulting $x_s\left(t\right)$ is plotted in Fig.~\ref{fig:x_shock}. In contrast to the one-point one-time problem \citep{MKV} the shock velocity $\dot{x}_{s}\left(t\right)$ here is not constant: it slowly decreases with time.  At $t\ll 1$ we obtain $x_s\left(t\right)\simeq \sqrt{-2\bar{H}} \, t$; the corresponding shock speed is equal to the one half of $V\left(x,t\right)$ at $t\ll 1$ [see Eq.~(\ref{eq:V_joint_region})], as it should \cite{Whitham}. At $t\to 1\;$ $x_s\left(t\right)$ approaches the point $x_s\left(t=1\right)=2^{1/6} \sqrt{-\bar{H}}$.

Notice that the solution~(\ref{eq:hsol_Hopf}) does not satisfy the flat initial condition~(\ref{eq:flat_IC}), but this is of no concern because $x_{s}\left(t=0\right)=0$, so the nontrivial Hopf region~(\ref{eq:nontrivial_Hopf_region}) is nonexistent at $t=0$.
At $t=0$ the entire system is in the trivial region where $h$ vanishes.

Our numerical results show good agreement with the analytic predictions, see Fig.~\ref{fig:negative_tail}. For example, for $\bar{H} = -50$ the action computed on the numerical solution, $s \simeq 2647$, is about $1\%$ off the analytical prediction, $s = 2666.66\dots$ of Eq.~(\ref{eq:action_negative_tail}).

The adiabatic soliton theory presented here can be extended to a more general problem of finding the probability distribution of observing a given height history at $x=0$, see Eq.~(\ref{moregeneral}). We determine the corresponding tail of this probability distribution, in Sec. \ref{sec:general_soliton}. We also show there how the solution of the more general problem can be used to reproduce some of the results of this subsection.

\section{$\lambda \bar{H} \to -\infty$ tail:  How a negative-pressure gas leaks into a wall}
\label{sec:HD}

In this tail of the distribution $\mathcal{P}\left(\bar{H},T\right)$, the optimal path is large-scale in terms of both $h$ and $\rho$, and we can neglect the diffusion terms
in Eqs.~(\ref{eqh}) and (\ref{eqrho}) altogether \cite{KK2007,MKV}. The resulting equations
\begin{eqnarray}
\label{eqV_HD}
\partial_{t}V+V\partial_{x}V&=&\partial_{x}\rho,\\
\label{eqrho_HD}
\partial_{t}\rho+\partial_{x}\left(\rho V\right)&=&-\Lambda\delta\left(x\right),
\end{eqnarray}
describe a one-dimensional inviscid hydrodynamics (HD) of a compressible ``gas" with density $\rho\left(x,t\right)$ and velocity $V\left(x,t\right) =\partial_x h \left(x,t\right)$ \cite{MKV}.
The gas is unusual: it has negative pressure $p\left(\rho\right)=-\rho^2/2$.
As described by Eqs.~(\ref{eq:flat_IC}) and~(\ref{pT}), the gas flow starts from rest, $V=0$ at time $t=0$, and it leaks, at constant rate $\Lambda$, into the origin until time $t=1$ when no gas is left.

\begin{figure}
\includegraphics[width=0.445\textwidth,clip=]{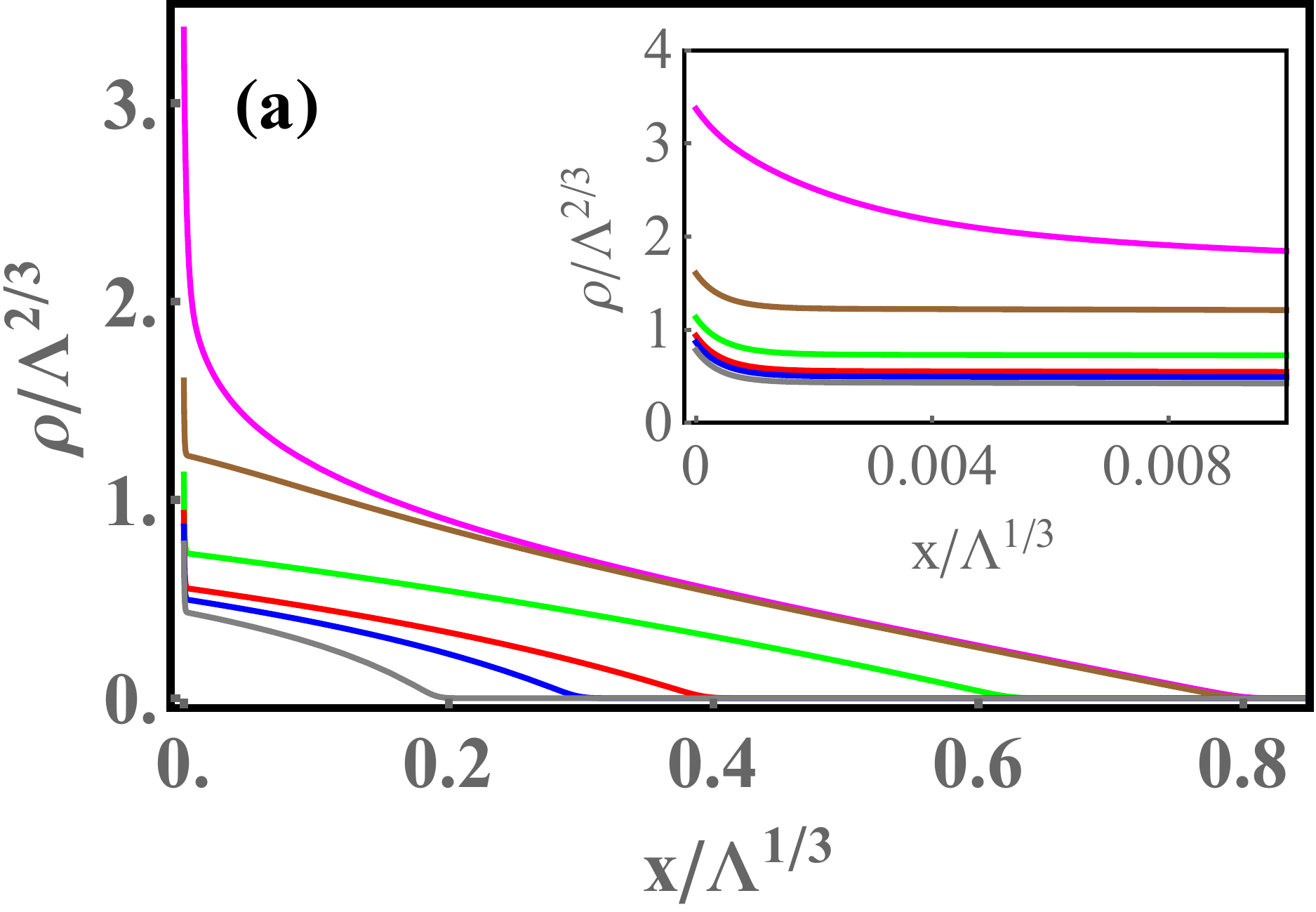}
\includegraphics[width=0.45\textwidth,clip=]{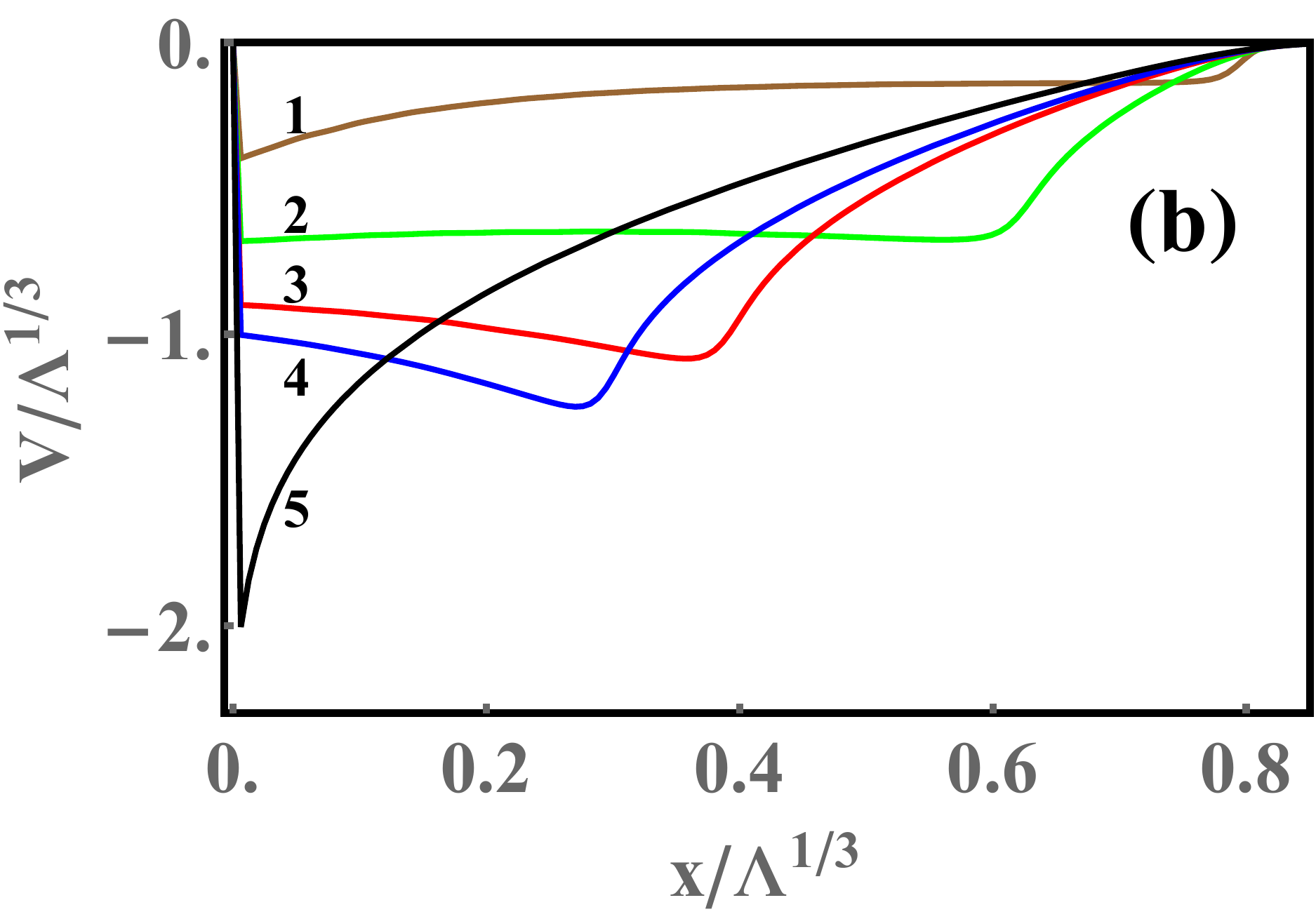}
\includegraphics[width=0.455\textwidth,clip=]{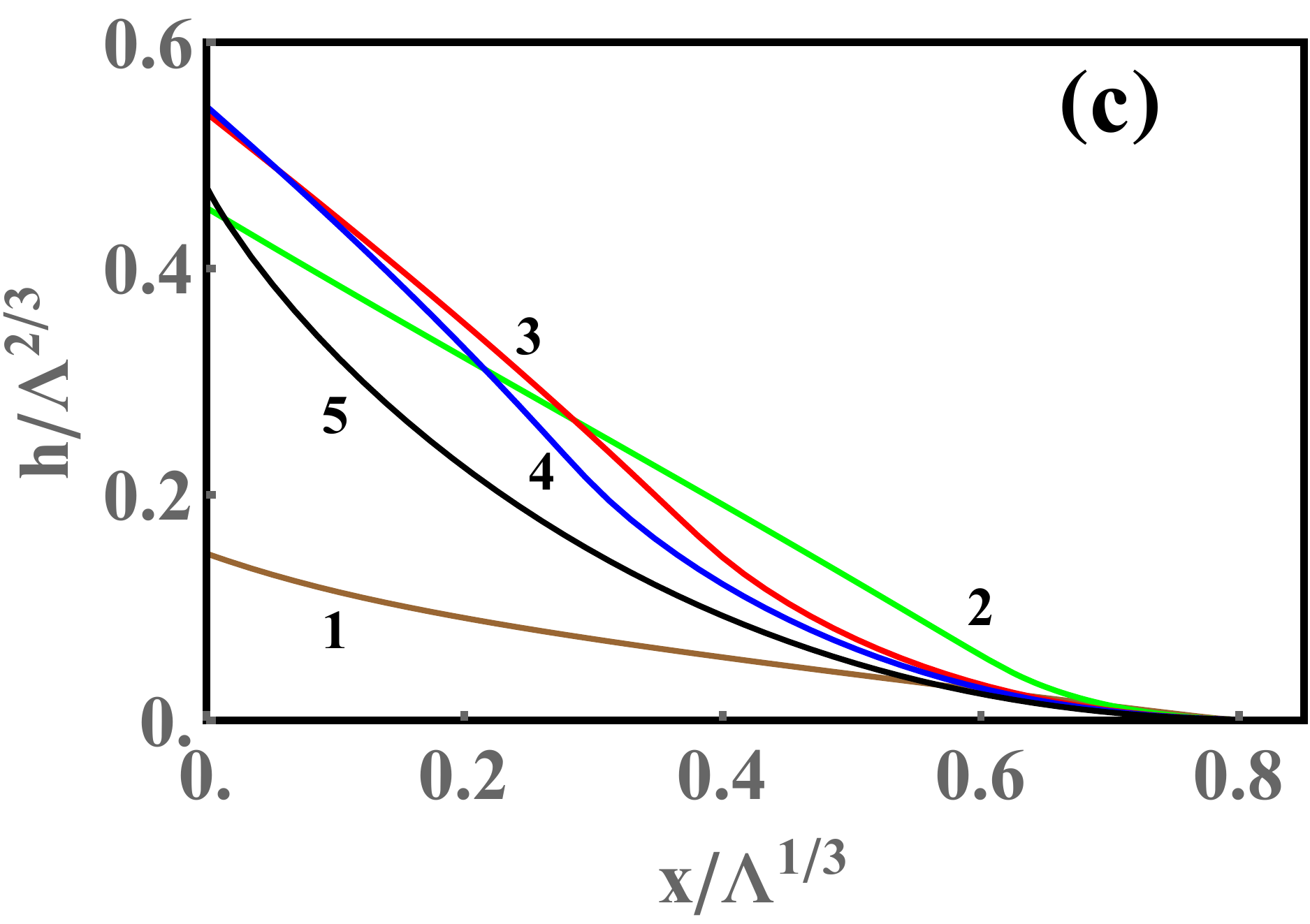}
\includegraphics[width=0.45\textwidth,clip=]{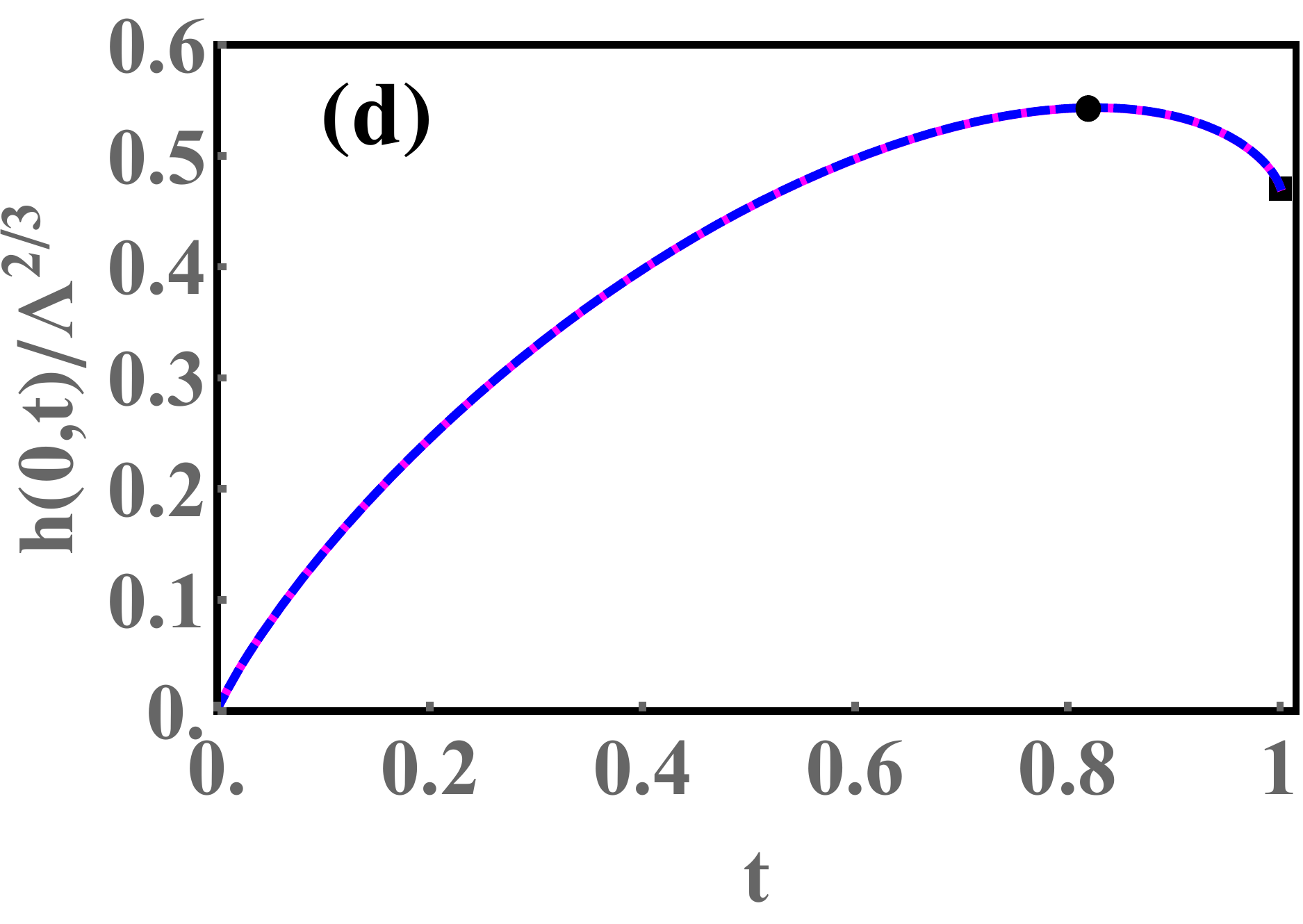}
\caption{Numerical results for the optimal path (at $x>0$), corresponding to the $-\lambda \bar{H} \gg 1$ tail. (a) Rescaled $\rho\left(x,t\right)$ at times $t =0,  0.1,  0.5,  0.75, 0.82$ and  $0.9$ (top to bottom). Inset: the diffusion-dominated boundary layer near $x=0$. (b) Rescaled $V\left(x,t\right)$ at times $0.1 (1),\,0.5 (2),\,0.75 (3),\,0.82 (4)$ and $1 (5)$. (c) Rescaled $h\left(x,t\right)$ at the same times as in (b). (d) Rescaled $h\left(x=0,t\right)$ vs. $t$. The coordinates of indicated points are (0.826, 0.543) --- the maximum height --- and (1, 0.472).
All the data is plotted for two different values of $\Lambda$: $75700$ and $97500$, which correspond to $\bar{H} \simeq 704,\,s \simeq 2.14\times10^{7}$ and $\bar{H}\simeq834,\,s\simeq3.26\times10^{7}$, respectively. The two data sets are indistinguishable, verifying the hydrodynamic scaling (\ref{HDrescaling}). The pressure-driven flow region and the Hopf region are clearly seen in panels (a)-(c).
The regime $x<0$ is given by symmetrization of $\rho$ and $h$ and by antisymmetrization of $V$. }
\label{fig:rho_and_V_positive_tail}
\end{figure}

The additional HD rescaling
\begin{equation}\label{HDrescaling}
 x/\Lambda^{1/3}\to x,\quad V/\Lambda^{1/3}\to V,\quad\rho/\Lambda^{2/3}\to\rho,
\end{equation}
leaves Eqs.~(\ref{eqV_HD}), (\ref{eq:flat_IC}) and~(\ref{pT}) invariant and replaces $\Lambda$ by $1$ in Eq.~(\ref{eqrho_HD}). As a result, the HD problem becomes parameter-free, and one immediately obtains the scalings $s\left(\Lambda\right)\sim\Lambda^{5/3}$ and $\bar{H}\left(\Lambda\right)\sim\Lambda^{3/2}$ leading to
\begin{equation}
\label{s_h_five_halves}
s\left(\bar{H}\right)=s_{0}\bar{H}^{5/2},
\end{equation}
which corresponds to Eq.~(\ref{5/2tail})
in the physical units. Here
$s_{0}=O(1)$ is a dimensionless number to be found. Another way to obtain the scaling~(\ref{5/2tail}) is to require the distribution~(\ref{actiondgen}) to be independent of the diffusion coefficient $\nu$.

To complete the formulation of the HD problem, we derive an effective boundary condition at $x=0$. Due to the mirror symmetry of the problem, $x \leftrightarrow -x$,
the solution should satisfy the relations $\rho\left(-x,t\right)=\rho\left(x,t\right)$ and $V\left(-x,t\right)=-V\left(x,t\right)$. As a result, $\rho\left(x,t\right)$ of the HD solution must be continuous at $x=0$, see also Eq.~(\ref{eqrho}). Integrating Eq.~(\ref{eqrho_HD}) (with $\Lambda=1$) with respect to $x$ over an infinitesimally small interval which includes $x=0$, we find that the HD solution must include a $V$-shock at $x=0$ which satisfies the relation
\begin{equation}
\left.\rho V\right|_{x\to0^{+}}-\left.\rho V\right|_{x\to0^{-}}=-1.
\end{equation}
Therefore, it suffices to solve Eqs.~(\ref{eqV_HD}) and~(\ref{eqrho_HD}) for $x>0$: without the delta-function term in Eq.~(\ref{eqrho_HD}), but with the following effective boundary condition:
\begin{equation}
\label{eq:constant_flux}
\left.\rho V\right|_{x\to0^{+}}=-\frac{1}{2}.
\end{equation}
This boundary condition describes a constant mass flux of the gas into the origin.
Similarly to the previous works \citep{MKV}, this HD flow has three distinct regions: (a) the pressure-driven flow region $0<x<\ell\left(t\right)$ with an a priori unknown $\ell\left(t\right)$, (b) the Hopf region $\ell\left(t\right) < x < \ell_0$, where $\ell_0=\ell\left(t=0\right)$, and (c) the trivial region $x \ge \ell_0$ where $\rho$ and $V$ vanish identically. In the Hopf region $\rho\left(x,t\right)$ vanishes identically, and Eq.~(\ref{eqV_HD}) becomes the Hopf equation~(\ref{eq:Hopf}) whose general solution~(\ref{eq:Hopf_gen_sol}) is to be matched continuously with the pressure-driven solution at $x = \ell\left(t\right)$ and must vanish at $x=\ell_0$.

In spite of the major simplification, provided by the inviscid approximation, we have not been able to solve the HD problem analytically. The reason is the nonzero-flux boundary condition~(\ref{eq:constant_flux}). (For the zero mass flux at the origin the HD solution describes a
uniform-strain flow, and it is quite simple \cite{KK2009,MKV}.) Therefore, we relied on a numerical solution.

\begin{figure}
\includegraphics[width=0.41\textwidth,clip=]{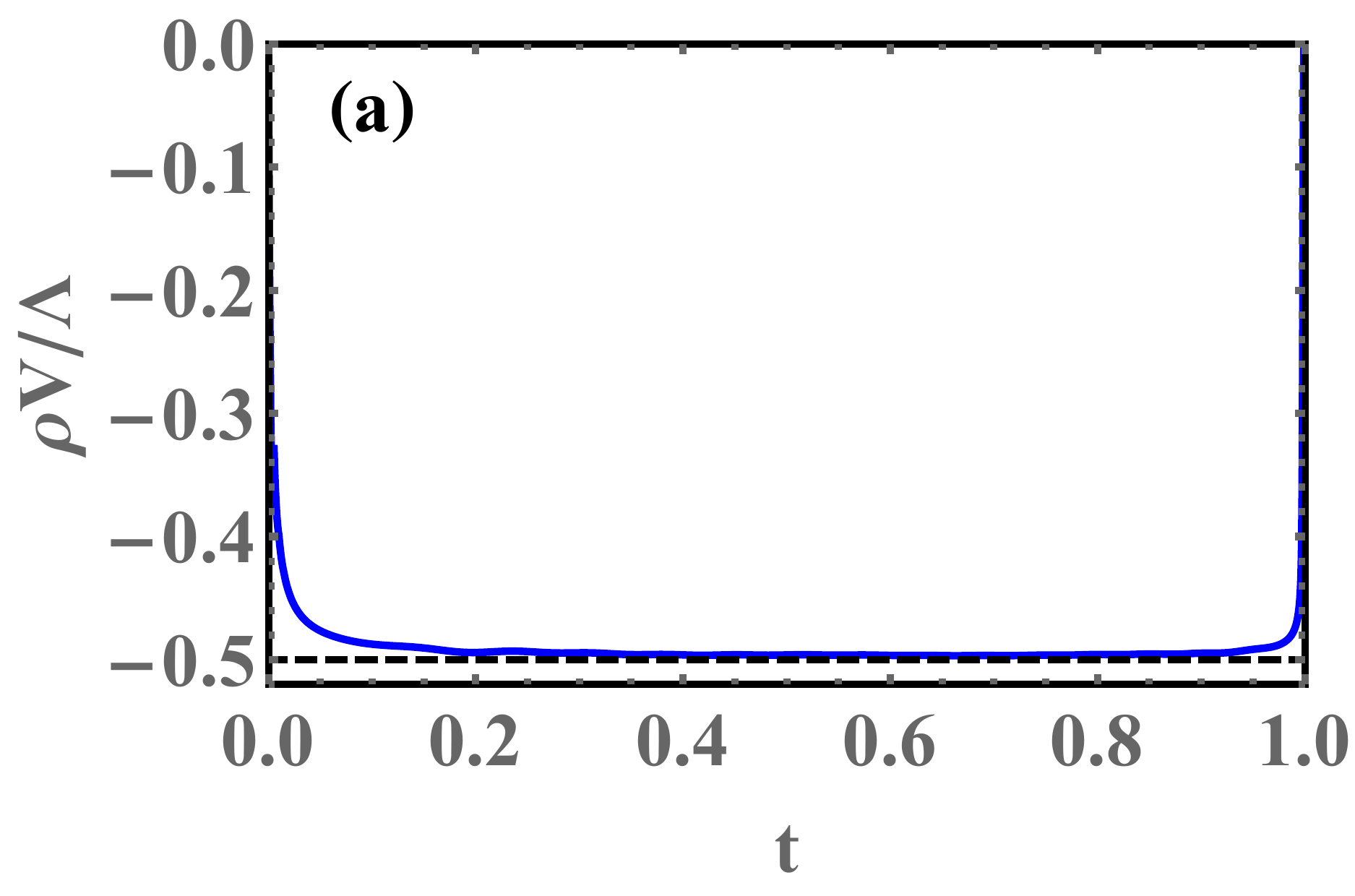}
\includegraphics[width=0.41\textwidth,clip=]{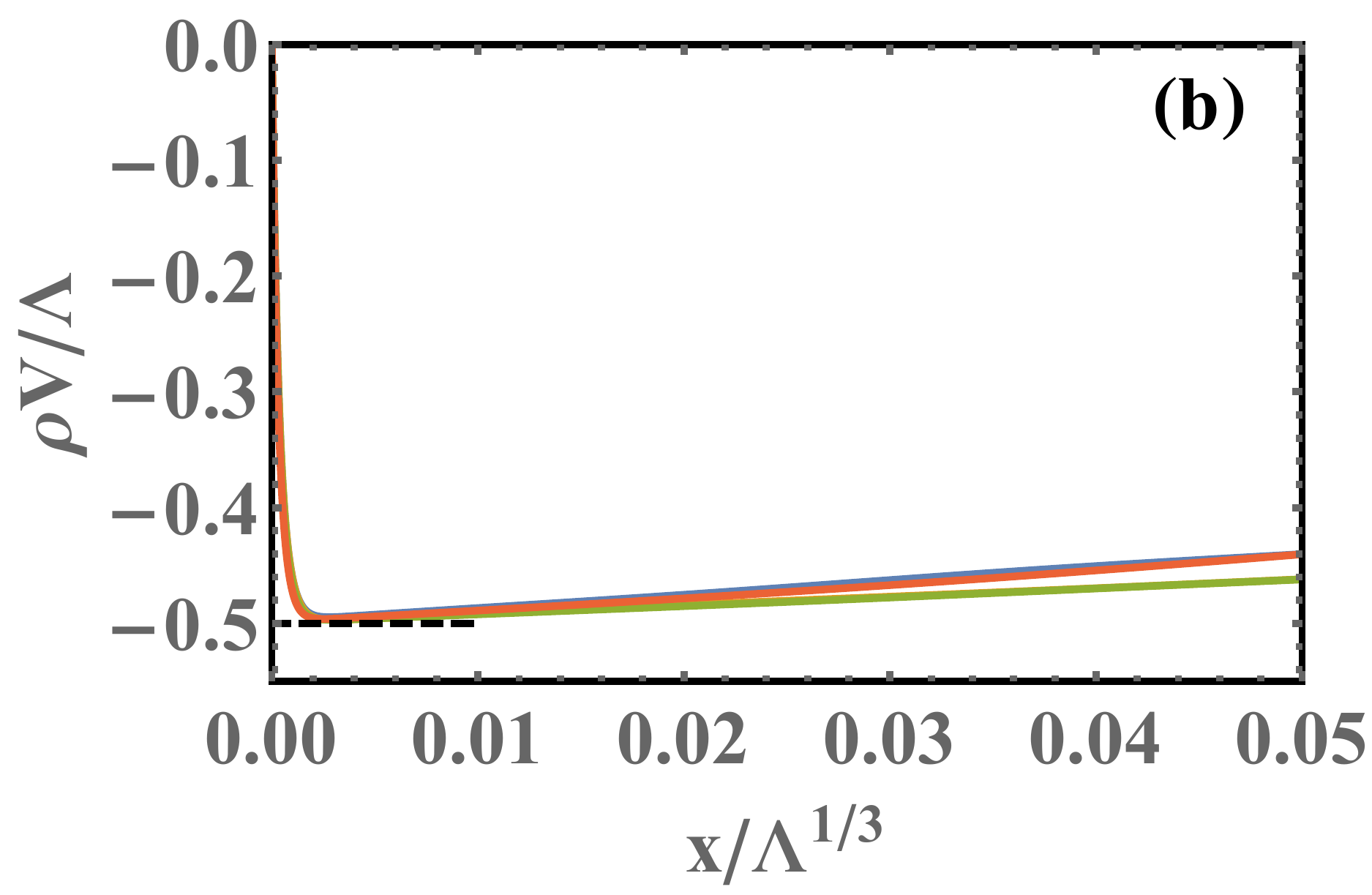}
\caption{(a) The rescaled ``mass flux" $\rho V$ at $x=0.2$ as a function of $t$. (b) $\rho V$ as a function of rescaled $x$ at times $t=0.25, 0.5, 0.75,0.9$.
Both plots are for $\Lambda = 97500$.
The effective HD boundary condition~(\ref{eq:constant_flux}) is satisfied except in narrow temporal boundary layers near $t=0$ and $t=1$, and in a narrow spatial boundary layer around $x=0$.}
\label{fig:flux_positive_tail}
\end{figure}

Numerical solutions for one-dimensional inviscid gas flows are usually obtained in Lagrangian mass coordinates \cite{ZR}. In the context of the OFM theory for the KPZ equation it been recently done in Ref. \cite{Asida2019} which studied the one-time statistics of the interface height at an arbitrary point on a half-line. In the present problem the use of the Lagrangian mass coordinate is inconvenient because of the ``mass leakage" at $x=0$. Therefore, we solved numerically the full OFM equations~(\ref{eqh}) and~(\ref{eqrho})
in the regime $\bar{H} \gg 1$. The numerical solutions verified the HD scaling (\ref{HDrescaling}), see Fig.~\ref{fig:rho_and_V_positive_tail}. As can be seen in panel (b), the $V$-shock, predicted by the HD equations at $x=0$, is smoothed out by diffusion over a narrow boundary layer. The pressure-driven flow region and the Hopf region are also clearly seen in Fig.~\ref{fig:rho_and_V_positive_tail}.  We also checked that the effective HD boundary condition~(\ref{eq:constant_flux}) holds, see Fig.~\ref{fig:flux_positive_tail}. From Fig.~\ref{fig:rho_and_V_positive_tail} (b) it can be seen that $\ell_0 \simeq 0.8$.

\begin{figure}
\includegraphics[width=0.43\textwidth,clip=]{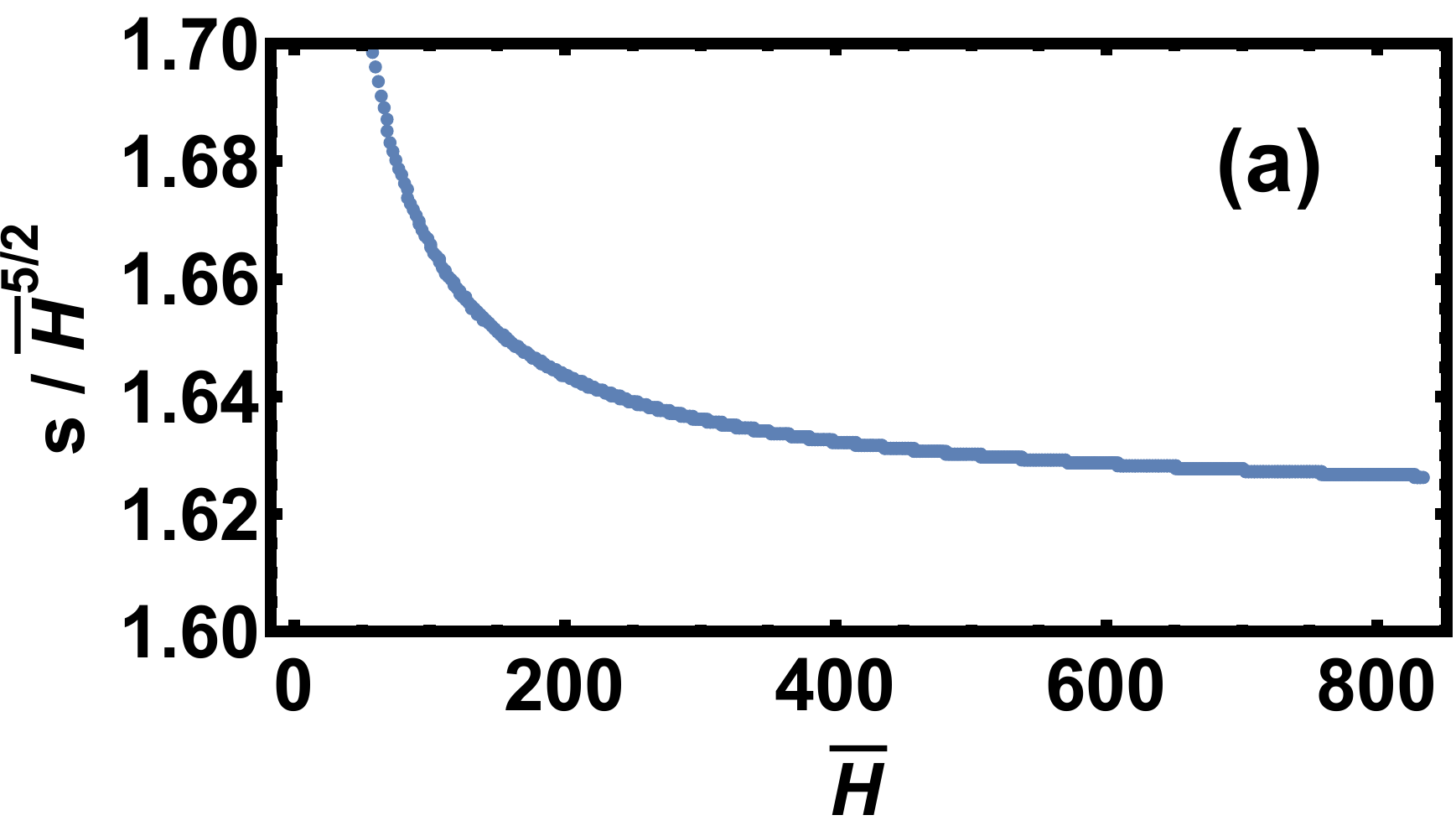}
\includegraphics[width=0.46\textwidth,clip=]{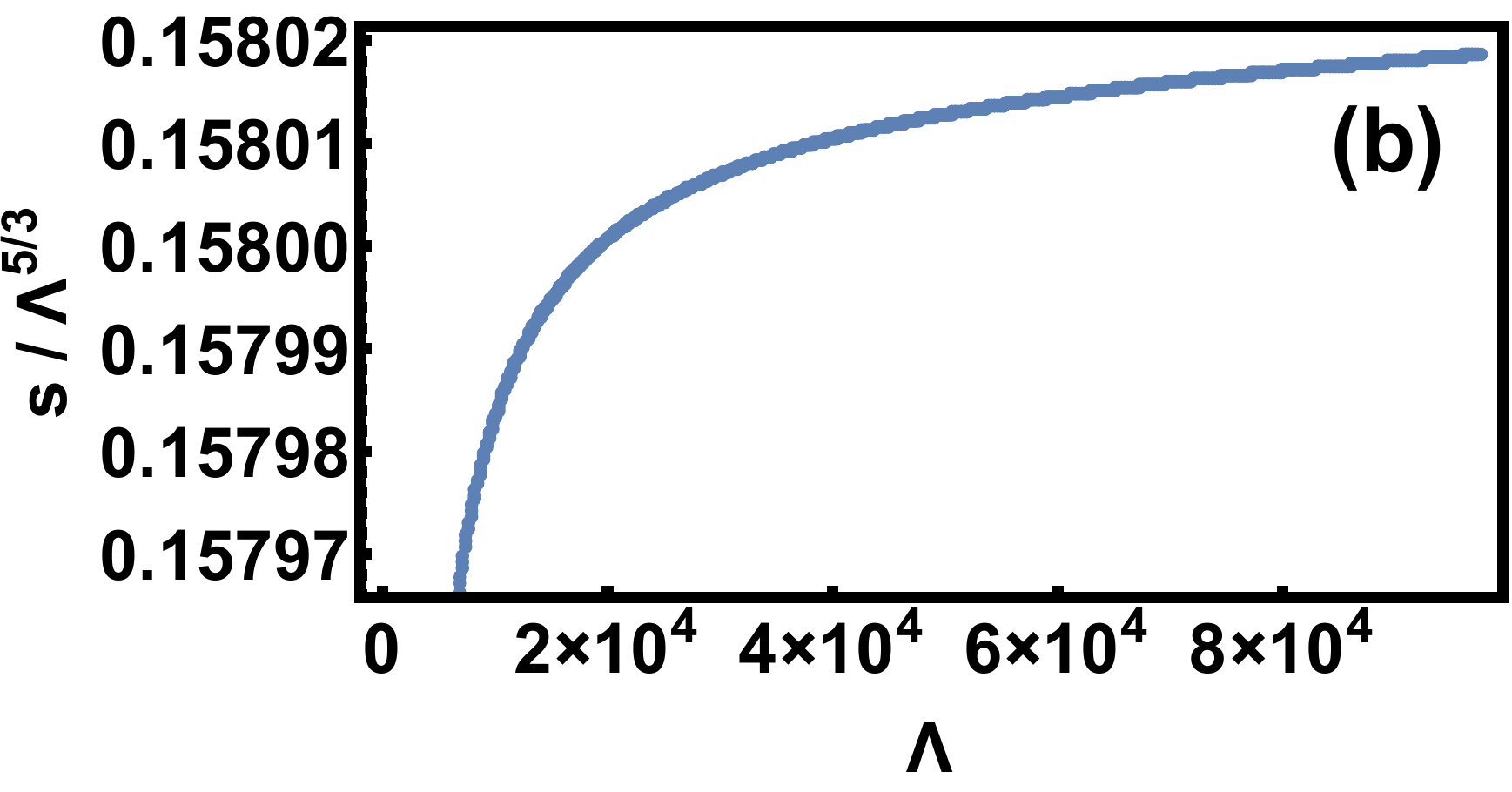}
\caption{The action computed numerically in the $-\lambda \bar{H} \gg 1$ tail. (a) $s\left(\bar{H}\right)/\bar{H}^{5/2}$ as a function of $\bar{H}$. (b) $s(\Lambda)/\Lambda^{5/3}$ as a function of $\Lambda$.
Both graphs approach constants in the limit $\bar{H}, \Lambda \to +\infty$. However, as it is evident from the large difference between the vertical scales, $s(\Lambda)/\Lambda^{5/3}$ converges much faster.
}
\label{fig:S_vs_H_positive_tail}
\end{figure}

Most importantly, we verified the scaling relation~(\ref{s_h_five_halves}) of the action, see Fig.~\ref{fig:S_vs_H_positive_tail}. Using the numerical data, we computed the factor $s_{0}=O(1)$ in the following manner.
The relation%
\footnote{See footnote \ref{footnote_ds_dH_Lambda}.}
$ds/d\bar{H}=\Lambda$ alongside with Eq.~(\ref{s_h_five_halves}) leads to
\begin{equation}
\label{eq:s_vs_lambda_positive_tail}
s\left(\Lambda\right)\simeq\left(\frac{2}{5}\right)^{5/3}s_{0}^{-2/3}\Lambda^{5/3},\quad\Lambda\gg1.
\end{equation}
As seen in Fig.~\ref{fig:S_vs_H_positive_tail}, $s\left(\bar{H}\right)/\bar{H}^{5/2}$ as a function of $\bar{H}$ converges relatively slowly in the $\bar{H}\to+\infty$ tail (a similar slow convergence of $s\left(\bar{H}\right)/\bar{H}^{5/2}$ for the one-point statistics was observed in the one-point one-time problem, see the inset of Fig.~1 of Ref.~\cite{MKV}).
Fortunately,  $s(\Lambda)/\Lambda^{5/3}$ as a function of $\Lambda$ converges much faster, see Fig.~\ref{fig:S_vs_H_positive_tail} (b).  By comparing Eq.~(\ref{eq:s_vs_lambda_positive_tail}) with the data of Fig.~\ref{fig:S_vs_H_positive_tail} (b), we find $\left(2/5\right)^{5/3}s_{0}^{-2/3}\simeq0.158$. The resulting $s_0 \simeq 1.61$ turns out to larger by a factor of $6.7$ than the corresponding factor $8\sqrt{2}/(15\pi)$ for the one-point one-time height statistics tail $\lambda H \to -\infty$ \cite{KK2009,MKV}.

\section{Adiabatic soliton theory for a more general problem}
\label{sec:general_soliton}

Here we consider the following problem: What is the probability density $\mathcal{P}\left[h_{0}\left(t\right)\right]$ of observing a whole given one-point height history
\begin{equation}
h\left(x=0,t\right)=h_{0}\left(t\right)
\end{equation}
of the KPZ interface at the origin?
The OFM formulation of this problem is a generalization of that of the time-average problem. We argue that the only difference is that the second OFM equation~(\ref{eqrho}) must be replaced by
\begin{equation}
\label{eqrho_generalized}
\partial_{t}\rho=-\partial_{x}^{2}\rho-\partial_{x}\left(\rho\partial_{x}h\right)-\Lambda_{0}\left(t\right)\delta\left(x\right),
\end{equation}
where $\Lambda_{0}\left(t\right)$ is a function which takes the role of (infinitely many) Lagrange multipliers, and is ultimately determined by the specified $h_{0}\left(t\right)$.
One way to reach Eq.~(\ref{eqrho_generalized}) is by imposing a finite number of intermediate-time constraints
\begin{equation}
h\left(0,t_{i}\right)=h_{0,i},\quad i=1,\dots,N
\end{equation}
at times $0<t_{1}<\dots<t_{N}<1$, and then taking the continuum limit.

Let us focus on histories
\begin{equation}
h_{0}\left(t\right) = Mg\left(t\right),
\end{equation}
where $g\left(0\right)=0$, $ g\left(1\right)=1$ and $g\left(t\right)$ is monotone increasing, $\dot{g}\left(t\right)>0$, and consider the limit $M \to -\infty$ (the reason for these requirements will become clear shortly).
We argue that the adiabatic soliton ansatz~(\ref{eq:rho_adiabatic_soliton}) and~(\ref{eq:h_adiabatic_soliton}) is still correct.  $c\left(t\right)$ is found from the adiabatic relation~(\ref{eq:c_h0dot}), and is related to $\Lambda_{0}\left(t\right)$ through Eq.~(\ref{eq:Lambda0_and_c}) with the right-hand side replaced by $\Lambda_{0}\left(t\right)$.
We must require $c\gg1$ for the adiabatic soliton theory to be valid, and this is the reason that we demanded that $g\left(t\right)$ be monotonic and considered $M \to -\infty$.
The action associated with the distribution $\mathcal{P}\left[h_{0}\left(t\right)\right]$ is given by a simple generalization of Eq.~(\ref{eq:action_negative_tail}), which is obtained by using Eq.~(\ref{eq:c_h0dot}) instead of~(\ref{eq:c_of_t_for_time_average}):
\begin{equation}
\label{eq:action_as_function_of_h0}
s=\frac{8\sqrt{2}}{3}\int_{0}^{1}\left[-\frac{dh_{0}(t)}{dt}\right]^{3/2}dt.
\end{equation}
Here too the outer region, where $\rho\left(x,t\right)\simeq 0$, does not contribute to the action and does not affect this distribution tail. As a result, the initial condition $h\left(x,t=0\right)$ does not play a major role, and
this tail is universal for a whole class of deterministic initial conditions.

Equation~(\ref{eq:action_as_function_of_h0}) can be used to find the corresponding tail of many simple KPZ height statistics that can be viewed as particular examples. We now demonstrate this by reproducing the tails of the distributions of the one-point one-time height $H$ \cite{KK2007,MKV} and of the time-averaged height $\bar{H}$.

We obtain the $H \to -\infty$ tail of the one-point one-time height distribution by a minimization of the action~(\ref{eq:action_as_function_of_h0}) over histories $h_{0}\left(t\right)$ at the origin. The ensuing Euler-Lagrange equation reduces to $d^{2}h_{0}/dt^{2}=0$, and its solution subject to the boundary conditions $h_{0}\left(0\right)=0$ and $h_{0}\left(1\right)=H$ is $h_{0}\left(t\right)=Ht$. Plugging $h_{0}\left(t\right)$ back into~(\ref{eq:action_as_function_of_h0}), we obtain the tail
\begin{equation}\label{onetime}
s=\frac{8\sqrt{2}}{3}\left|H\right|^{3/2}
\end{equation}
in agreement with Refs.~\cite{KK2007,MKV}.

For the $\bar{H}\to -\infty$ tail of the time-averaged height distribution we must minimize~(\ref{eq:action_as_function_of_h0}) over $h_{0}\left(t\right)$, under the constraints $h_{0}\left(0\right)=0$ and Eq.~(\ref{eq:hbar_def}). The constraint~(\ref{eq:hbar_def}) calls for a Lagrange multiplier, so we define
\begin{equation}
\label{eq:s_Lambda_def_h0}
s_{\Lambda}=\frac{8\sqrt{2}}{3}\int_{0}^{1}\left[-\frac{dh_{0}(t)}{dt}\right]^{3/2}dt-\Lambda\int_{0}^{1}h_{0}\left(t\right)dt,
\end{equation}
while the ``lacking" boundary condition is the condition $d{h}_{0}/dt(t=1) =0$ at the  ``free boundary" $t=1$ \cite{Elsgolts}. The Euler-Lagrange equation associated with $s_{\Lambda}$,
\begin{equation}
\frac{d^{2}h_{0}(t)}{dt^{2}}=-\frac{\Lambda}{2\sqrt{2}}\left[-\frac{dh_{0}(t)}{dt}\right]^{1/2},
\end{equation}
is equivalent to both of the equations~(\ref{eq:Lambda0_and_c}) and~(\ref{eq:c_h0dot}), and its solution under the constraints listed above indeed coincides with Eq.~(\ref{eq:h0_sol}). Plugging Eq.~(\ref{eq:h0_sol}) back into~(\ref{eq:action_as_function_of_h0}) this indeed reproduces the tail~(\ref{eq:action_negative_tail}).

One can also use Eq.~(\ref{eq:action_as_function_of_h0}) to obtain a proper tail of the joint distribution of $H$ and $\bar{H}$.

\section{Summary and discussion}
\label{disc}

We calculated the Gaussian asymptote~(\ref{gauss}), the third cumulant~(\ref{eq:kappa3}) and the two stretched exponential tails~(\ref{3/2tail}) and~(\ref{5/2tail}) of the short-time distribution of the time-averaged height $\bar{H}$ at a given point of an infinite  initially flat KPZ interface in 1+1 dimensions. We also found the corresponding optimal path of the interface, conditioned on a given $\bar{H}$. The scaling of the logarithm of the distribution with $\bar{H}$ turns out to be the same as that observed for one-point statistics, but the details (and especially the optimal paths) are quite different. In all regimes we observed that  the probability to observe a certain unusual value of $\bar{H}$ is smaller than the probability to observe the same value of $H$ in the one-time problem. The optimal fluctuation method makes it obvious why: in order to reach a given $\bar{H}$, the optimal interface height must reach a higher value at an earlier time.

One non-intuitive feature that we observed is a non-monotonic behavior of the optimal interface height at $x=0$ as a function of time. This effect is most pronounced for the typical fluctuations of $\bar{H}$ (as described by the Gaussian region of the distribution) and in the $\lambda \bar{H} \to -\infty$ tail. It is much weaker in the $\lambda \bar{H} \to +\infty$ tail.

Solving the OFM equations analytically in the $\lambda \bar{H} \to -\infty$ tail remains challenging even after a drastic reduction of the problem to that of an effective hydrodynamic flow, as we described in Sec. \ref{sec:HD}.

It would be interesting to extend our results to the other two standard initial conditions of the KPZ equation: droplet and stationary. We expect the short-time scaling of the distribution to be the same.  For the droplet initial condition (actually,  for a whole class of deterministic initial conditions),  we expect the $\lambda \bar{H} \to +\infty$ tail to coincide with that for the flat initial condition. For the stationary interface it would be interesting to find out whether the dynamical phase transition, reflecting spontaneous breaking of the mirror symmetry by the optimal path and observed in the one-point one-time problem \citep{Janas2016, KrajenbrinkLeDoussal2017, SKM2018}, persists in the statistics of $\bar{H}$.

The adiabatic soliton theory of Sec.~\ref{sec:soliton} can be also very useful in determining other types of short-time height statistics. We demonstrated this in Sec.~\ref{sec:general_soliton} by applying this theory to the more general problem of calculating the probability density of observing a given height history at the origin.

Finally, it would be very interesting, but challenging,  to study the time-average height statistics in the long-time limit, or even at arbitrary times. In analogy with the one-point one-time distribution \citep{MeersonSchmidt2017,SMP,KrajenbrinkLD2018tail,Corwinetal2018,Krajenbrinketal2018}, it is reasonable to expect that the large-deviation distribution tails, predicted in this work,  will continue to hold, sufficiently far in the tails, at arbitrary times.

\section*{ACKNOWLEDGMENTS}

We thank Tal Agranov and Joachim Krug for useful discussions and acknowledge financial support from the Israel Science Foundation (grant No. 807/16). N.R.S. was supported by the Clore foundation.



\begin{appendices}
\section{Derivation of OFM equations and boundary conditions}
\label{appendix:OFMequations}

\renewcommand{\theequation}{A\arabic{equation}}
\setcounter{equation}{0}

The variation of the modified action~(\ref{eq:stotal_def}) is
\begin{equation}\label{variation}
\delta s_{\Lambda}=\int_{0}^{1}dt\int_{-\infty}^{\infty}dx\left[\partial_{t}h-\partial_{x}^{2}h+\frac{1}{2}\left(\partial_{x}h\right)^{2}\right]\left(\partial_{t}\delta h-\partial_{x}^{2}\delta h+\partial_{x}h\,\partial_{x}\delta h\right)-\Lambda\int_{0}^{1}dt\,\delta h\left(x=0,t\right).
\end{equation}
Let us introduce the momentum density field $\rho\left(x,t\right)=\delta L/\delta\left(\partial_{t}h\right)$, where
$$
L\left\{ h\right\} =\frac{1}{2}\int_{-\infty}^{\infty} dx \left[\partial_{t}h-\partial_{x}^{2}h+\frac{1}{2}\left(\partial_{x}h\right)^{2}\right]^{2}
$$
is the Lagrangian.  We obtain
\begin{equation}\label{p}
\rho=\partial_{t}h-\partial_{x}^{2}h+\frac{1}{2}\left(\partial_{x}h\right)^{2},
\end{equation}
which can be rewritten as Eq.~(\ref{eqh}), the first Hamilton equation of the OFM. Now we can rewrite the variation (\ref{variation}) as follows:
\begin{equation}
\delta s_{\Lambda}=\int_{0}^{1}dt\int_{-\infty}^{\infty}dx\,\rho\,(\partial_{t}\delta h-\partial_{x}^{2}\delta h+\partial_{x}h\,\partial_{x}\delta h)-\Lambda\int_{0}^{1}dt\int_{-\infty}^{\infty}dx\,\delta h\left(x,t\right)\,\delta\left(x\right).
\end{equation}
Requiring the variation to vanish for arbitrary $\delta h$ yields, after several integrations by parts, the second Hamilton equation~(\ref{eqrho}) of the OFM.
The boundary terms in space, resulting from the integrations by parts, all vanish.
The boundary terms in time must vanish independently at $t=0$ and $t=1$. They vanish at $t=0$ because of the deterministic initial condition~(\ref{eq:flat_IC}), and the boundary term at $t=1$ leads to the boundary condition~(\ref{pT}).

\section{Lower cumulants}
\subsubsection{Optimal path in the Edwards-Wilkinson regime}
\label{appendix:h_EW_regime}

\renewcommand{\theequation}{B\arabic{equation}}
\setcounter{equation}{0}

The optimal path in the EW regime is obtained by solving Eq.~(\ref{heqEW}) with the initial condition (\ref{eq:flat_IC}) and with the forcing term $\rho\left(x,t\right)$ from Eq. (\ref{eq:rhoEW}). Using ``Mathematica" \citep{Mathematica}, we obtain
\begin{eqnarray}
\!\!\!h\left(x,t\right)&=&\frac{\Lambda}{24}\left\{ x\left[2x\left|x\right|+\left(6t-x^{2}-6\right)\text{erf}\left(\frac{x}{2\sqrt{1-t}}\right)-2\left(6t+x^{2}\right)\text{erf}\left(\frac{x}{2\sqrt{t}}\right)+\left(6t+x^{2}+6\right)\text{erf}\left(\frac{x}{2\sqrt{t+1}}\right)\right]\right.\nonumber\\
&+& \!\frac{2}{\sqrt{\pi}}e^{-\frac{(2t+1)x^{2}}{4t(t+1)}}\!\left.\left[\sqrt{1-t}\,e^{\frac{\left(1-3t^{2}\right)x^{2}}{4t-4t^{3}}}\left(4t-x^{2}-4\right)-2\sqrt{t}\,e^{\frac{x^{2}}{4t+4}}\left(4t+x^{2}\right)+\sqrt{t+1}\,e^{\frac{x^{2}}{4t}}\left(4t+x^{2}+4\right)\right]\right\} \!,
\end{eqnarray}
so $\partial_{x}h\left(x,t\right)$, which is useful for evaluating the third cumulant, is
\begin{eqnarray}
\partial_{x}h\left(x,t\right)&=&\frac{\Lambda}{4}\left\{ \left(t-\frac{x^{2}}{2}-1\right)\text{erf}\left(\frac{x}{2\sqrt{1-t}}\right)
-\left(2t+x^{2}\right)\!\text{erf}\left(\!\frac{x}{2\sqrt{t}}\!\right)\!
+\frac{2t+x^{2}\!+2}{2}\text{erf}\left(\!\frac{x}{2\sqrt{t+1}}\!\right)\!+\!x\left|x\right|\right.\nonumber\\
&+&\left.\frac{x}{\sqrt{\pi}}\left[\sqrt{t+1}\,e^{-\frac{x^{2}}{4(t+1)}}-\sqrt{1-t}\,e^{\frac{x^{2}}{4(t-1)}}
-2\sqrt{t}\,e^{-\frac{x^{2}}{4t}}\right]\right\} .\nonumber\\
\end{eqnarray}

\subsubsection{Shortcut for evaluating the third cumulant}
\label{appendix:shortcut_3rd_cumulant}

We denote by
\begin{equation}
s_{\text{EW}}\left[h\left(x,t\right)\right]=\frac{1}{2}\int_{0}^{1}dt\int_{-\infty}^{\infty}dx\left(\partial_{t}h
-\partial_{x}^{2}h\right)^{2}
\end{equation}
the dynamical action, corresponding to the Edwards-Wilkinson equation.
The minimum of $s_{\text{EW}}$ under the conditions~(\ref{eq:hbar_def}) and~(\ref{eq:flat_IC}) is given by $h\left(x,t\right) = \bar{H}h_{1}\left(x,t\right)$, so the variational derivative
$\delta s_{\text{EW}}/\delta h$ vanishes on this profile for any $\delta h\left(x,t\right)$ which satisfies the coditions $\delta h\left(x,0\right)=0$ and $\int_{0}^{1}\delta h\left(0,t\right)dt=0$. Since $h_2(x,t)$ in Eq.~(\ref{eq:h_perturbation}) satisfies these conditions, we find
\begin{equation}
\label{eq:cubic_term_vanishes}
s_{\text{EW}}\left[\bar{H}h_{1}\left(x,t\right)+\bar{H}^{2}h_{2}\left(x,t\right)+\mathcal{O}\left(\bar{H}^{3}\right)\right]= s_{\text{EW}}\left[\bar{H}h_{1}\left(x,t\right)\right]+\mathcal{O}\left(\bar{H}^{4}\right) =\bar{H}^{2}s_{1} + \mathcal{O}\left(\bar{H}^{4}\right),
\end{equation}
that is, the term cubic in $\bar{H}$ vanishes.
The action~(\ref{eq:sdyn_def}) can be rewritten as
\begin{equation}
\label{eq:action_separated}
s=s_{\text{EW}}+\frac{1}{2} \! \int_{0}^{1} \!\! dt \! \int_{-\infty}^{\infty}\!\!\!\!dx\left[\left(\partial_{t}h-\partial_{x}^{2}h\right)\left(\partial_{x}h\right)^{2} \! +\frac{1}{4}\left(\partial_{x}h\right)^{4}\right].
\end{equation}
We now plug the perturbative expansion~(\ref{eq:h_perturbation}) into~(\ref{eq:action_separated}) and using Eq.~(\ref{eq:cubic_term_vanishes}) we obtain
\begin{equation}
 s=\bar{H}^{2}s_{1}+\frac{\bar{H}^{3}}{2}\!\int_{0}^{1}\!\!dt\!\int_{-\infty}^{\infty}\!\!\!\!dx\left(\partial_{t}h_{1}-\partial_{x}^{2}h_{1}\right)\left(\partial_{x}h_{1}\right)^{2}+\mathcal{O}\left(\bar{H}^{4}\right),
\end{equation}
So the cubic term in $s\left(\bar{H}\right)$ is $\bar{H}^{3}s_{2}$ where $s_2$ is given by Eq.~(\ref{eq:s2_shortcut}),
and we used the fact that $h_1(x,t)$ and $\rho_1(x,t)$ satisfy Eq.~(\ref{heqEW}).

\end{appendices}

\end{document}